\documentclass[12pt,a4paper]{JHEP3}
\usepackage{amsmath,amssymb,bm}

\def\Eqref#1{(\ref{#1})}

\def\Eq#1{\begin{equation} #1 \end{equation}}

\def\Eqr#1{\begin{eqnarray} #1 \end{eqnarray}}

\def\Eqrsubl#1#2{\begin{subequations}\label{#1}\Eqr{#2}\end{subequations}}
\def\Bitm{\begin{itemize}}
\def\Eitm{\end{itemize}}
\def\Blist#1#2{\begin{list}{#1}{\parsep=0pt \itemsep=0pt%
  \listparindent=0pt #2}}
\def\Elist{\end{list}}
\long\def\ignore#1#2{\def\ignoreflag{#1}\long\def\tmptext{#2}
  \ifnum\ignoreflag>1 #2 \fi}
\numberwithin{equation}{section}

\newcommand{\nn}{\nonumber\\}
\newcommand{\pd}{\partial}

\newcommand{\bea}{\begin{eqnarray}}
\newcommand{\ena}{\end{eqnarray}}
\newcommand{\pa}{\partial}
\renewcommand{\a}{\alpha}
\renewcommand{\b}{\beta}

\def\Xsp{{\rm X}}

\def\Ysp{{\rm Y}}

\def\Zsp{{\rm Z}}
\def\lap{{\triangle}}

\newcommand{\vs}[1]{\vspace{#1 mm}}

\numberwithin{equation}{section}

\title{Dynamics of intersecting brane systems \\
-- Classification and their applications --}
\author{
Kei-ichi Maeda\\
Department of Physics, Waseda University\\
~~3-4-1 Okubo, Shinjuku, Tokyo 169-8555, Japan.\\
Waseda Research Institute for Science and Engineering \\
~~3-4-1 Okubo, Shinjuku, Tokyo 169-8555, Japan.\\
~~~E-mail: \email{maeda$``$at$"$waseda.jp}}
\author{
Nobuyoshi Ohta\\
Department of Physics, Kinki University\\
~~Higashi-Osaka, Osaka 577-8502, Japan.\\
~~~E-mail: \email{ohtan$``$at$"$phys.kindai.ac.jp}}
\author{Kunihito Uzawa\\
Yukawa Institute for Theoretical Physics\\
~~Kyoto University, Kyoto 606-8502, Japan.\\
~~~E-mail: \email{uzawa$``$at$"$yukawa.kyoto-u.ac.jp},\\
Osaka City University Advanced Mathematical Institute\\
~~Sumiyoshi-ku, Osaka-shi 558-8585, Japan.\\
~~~E-mail: \email{uzawa$``$at$"$sci.osaka-cu.ac.jp}}

\abstract{%
We present dynamical intersecting brane solutions
in higher-dimensional gravitational theory coupled to dilaton and several forms.
Assuming the forms of metric, form fields, and dilaton field,
we give a complete classification of dynamical intersecting brane solutions
with/without M-waves and Kaluza-Klein monopoles in
eleven-dimensional supergravity.
We apply these solutions to cosmology and black holes.
It is shown that these give FRW cosmological solutions and
in some cases Lorentz invariance is broken in our world.
If we regard the bulk space as our universe, we may interpret
them as black holes in the expanding universe.
We also discuss lower-dimensional effective theories
and point out naive effective theories may
give us some solutions which are inconsistent with
the higher-dimensional Einstein equations.
}
\keywords{Time-dependent intersecting branes, Brane cosmology,
A time-dependent black hole, Lower-dimensional effective theory}
\preprint{KU-TP 030 \hfill}

\begin{document}
\section{Introduction}
 \label{sec:introduction}

Recently there have been works on dynamical spacetime-dependent solutions
of supergravity involving branes which are of cosmological interest.
The dynamical solutions of supergravity have a number of important
applications. In the original version~\cite{Gibbons:2005rt},
one considers a spacetime-dependent brane solution with five-form flux
and gravity in the ten-dimensional type IIB supergravity.
In the presence of the spacetime dependence in the background metric, one finds,
even for the general black $p$-brane system~\cite{Lu:1995cs,Lu2,Lu3}, that the
structure of warp factor which depends on the space and time is different from
the usual ``product type" ansatz~\cite{Duff:1986hr,Edelstein,Ohta:2006sw,Deger}.

In addition to spacetime-dependent brane solutions in higher-dimensional
supergravities, there are several analyses of lower-dimensional
effective theories after compactifying the internal
space~\cite{Kodama:2005cz, Koyama:2006ni, Shiu:2008ry, Douglas:2008jx}.
The same considerations also apply to the string theories which are
of much interest as an approach to behavior of the early universe.
However, it has been pointed out that the four-dimensional effective theories
for warped compactification of ten-dimensional type IIB supergravity
may not provide solutions in the original
higher-dimensional theories~\cite{Kodama:2005cz, Chen:2005jp}.
This caution can be generalized in
various $p$-brane solutions~\cite{Binetruy:2007tu}.

Another significant fact is that more general dynamical brane solutions
arise if the gravity is coupled not only to single gauge field but to
several combinations of scalars and forms as intersecting brane solutions
in the supergravity.
The intersecting brane solutions were originally found by
G\"{u}ven in eleven-dimensional supergravity~\cite{Gueven:1992hh}.
After that, many authors investigated related solutions such as
intersecting membranes, and they constructed static new solutions
of intersecting branes~\cite{Argurio:1997gt, Ohta:1997wp, Aref'eva:1997nz,Ohta:1997gw,
Youm:1999zs, Deger:2002ie,Ohta:2003uw, Ohta:2003rr, Miao:2004bn, Chen:2005uw}.
For a nice review, see \cite{Stelle:1998xg}.
Furthermore, a different class of dynamical brane solutions which depend on both
time and space coordinates have been found in~\cite{Kodama:2005fz}, and
special intersecting dynamical solutions of D4-D8 are given
in~\cite{Binetruy:2007tu}.

In the present paper, we give general dynamical solutions of
intersecting  brane systems in $D$-dimensional theories, which
may have more general applications to cosmology and black hole physics,
and discuss their implications to lower-dimensional effective theories.
We show that these solutions give FRW universe if we regard the homogeneous and
isotropic part of the brane world-volumes as our spacetime,
whereas they give black hole solutions in FRW universe if we regard the bulk transverse
space as our spacetime. We also show that in the former case, Lorentz invariance
may appear broken in our four-dimensional world though more elaborate
solutions may be necessary to obtain realistic models.
Although our solutions contain general intersecting brane solutions including
D-branes and NS-branes, we discuss M-branes mainly in our following
discussions for simplicity. Other branes can be obtained by dimensional reductions
and T-duality.

The paper is organised as follows.
In \S~\ref{sec:Dynamical intersecting p-branes solution},
we first consider intersecting $p$-brane systems
in $D$-dimensions and derive general dynamical intersecting brane
solutions under certain metric ans\"atze.
In \S~\ref{sec:classification}, focusing on intersecting M-brane systems
in the eleven-dimensional supergravity,
we give a classification of dynamical intersecting brane solutions without
M-wave and Kaluza-Klein(KK)-monopole, and discuss spacetime structure of
the intersecting branes. A complete classification of these solutions is given in
Appendix~\ref{Appendix:table_intersectingMbranes}.
In \S~\ref{application}, applications of these solutions to cosmology and
black hole physics are discussed.
In \S~\ref{Mwave_KKmonopole}, dynamical intersecting brane solutions
involving M-wave and KK-monopole are discussed together
with their applications to cosmology and black holes.
In \S~\ref{effective theory}, we discuss lower-dimensional effective theories
for the warped compactification of the brane systems in eleven-dimensional
supergravity and discuss that Lorentz invariance in our spacetime may appear
broken in our solutions.
\S~\ref{sec:Discussions} is devoted to concluding remarks.
Dynamical solutions of single branes are summarized in Appendix~\ref{Appendix:p},
and the complete classification of intersecting M-branes are given in
Appendix~\ref{Appendix:table_intersectingMbranes}.
Solutions with M-wave and KK-monopole are given in Appendix~\ref{KK_and_monopole},
and their intersections with M-branes are given in
Appendix~\ref{Intersecting_branes_with_MKK}.

\section{Solutions of dynamical intersecting branes}
\label{sec:Dynamical intersecting p-branes solution}

In this section, we consider dynamical intersecting brane systems
in $D$ dimensions. We write down the Einstein equations under certain
metric ans\"atze, which are a generalization of
those of known static intersecting $p$-brane solutions.
We then solve the Einstein equations and present the solutions explicitly.
To compare the results of intersecting $p$-brane with those of single
$p$-branes, we summarize the dynamical solutions
of single $p$-branes in Appendix~\ref{Appendix:p}.

Let us consider a gravitational theory with the metric $g_{MN}$, dilaton $\phi$,
and anti-symmetric tensor fields of rank $(p_I+2)$, where
$I$ denotes the type of the corresponding branes.
The most general action for the intersecting-brane system
is written as
\Eq{
S=\frac{1}{2\kappa^2}\int \left[R\ast{\bf 1}_D
 -\frac{1}{2}d\phi \wedge \ast d\phi
 -\sum_I \frac{1}{2(p_I+2)!}e^{c_I\phi}F_{(p_I+2)}
 \wedge\ast F_{(p_I+2)}\right],
\label{gs:D-dim action:Eq}
}
where $\kappa^2$ is the $D$-dimensional gravitational constant,
$\ast$ is the Hodge dual operator in the $D$-dimensional spacetime,
$c_I$ is a constant given by
\Eq
{
c_I^2=4-\frac{2(p_I+1)(D-p_I-3)}{D-2}.
\label{gs:parameter c:Eq}
}
The expectation values of fermionic fields are assumed to be zero.

The field equations are given by
\Eqrsubl{gs:field equations:Eq}{
&&R_{MN}=\frac{1}{2}\pd_M\phi \pd_N \phi
 +\frac{1}{2}\sum_I\frac{1}{(p_I+2)!}e^{\epsilon_Ic_I\phi}\nn
&&~~~~~~~~\times
\left[(p_I+2)F_{MA_2\cdots A_{p_I+2}} {F_N}^{A_2\cdots A_{p_I+2}}
-\frac{p_I+1}{D-2}g_{MN} F^2_{(p_I+2)}\right],
\label{gs:Einstein equations:Eq}\\
&&\Box\phi=\frac{1}{2}\sum_I\frac{\epsilon_Ic_I}{(p_I+2)!}
e^{c_I\phi}F^2_{(p_I+2)},
 \label{gs:scalar field equation:Eq}\\
&&d\left(e^{c_I\phi}\ast F_{(p_I+2)}\right)=0,
   \label{gs:gauge field equation:Eq}
}
where $\Box$ is the $D$-dimensional D'Alembertian.

To solve the field equations, we assume the $D$-dimensional metric of the form
\Eqr{
\hspace{-10mm}
ds^2={\cal A}(t, z)u_{ij}(z) dz^i dz^j-{\cal B}(t, z)dt^2
+\sum_{\alpha=1}^p{\cal C}^{(\alpha)}(t, z)
(dx^{\alpha})^2 ,
 \label{gs:metric:Eq}
}
where
$u_{ij}(z)$ is the metric of the $(D-p-1)$-dimensional $\Zsp$ space which
depends only on the $(D-p-1)$-dimensional coordinates $z^i$.
${\cal A}$, ${\cal B}$ and ${\cal C}^{(\alpha)}$ are given by
\Eq{
{\cal A}=\prod_I\left[h_I(t,z)\right]^{a_I},~~~
{\cal B}=\prod_I\left[h_I(t,z)\right]^{b_I},~~~
{\cal C}^{(\alpha)}=
\prod_I\left[h_I(t,z)\right]^{c^{(\alpha)}_I}.
}
where the parameters $a_I$, $b_I$ and $c^{(\alpha)}_I$ are defined by
\Eq{
a_I=\frac{p_I+1}{D-2},~~~~b_I=-\frac{D-p_I-3}{D-2},~~~~
c^{(\alpha)}_I=\left\{
\begin{array}{cc}
b_I&~{\rm for}~~\alpha\in I\\ a_I &~{\rm for}~~\alpha
\in \hspace{-.8em}/ I
\end{array} \right. ,
 \label{gs:paremeter:Eq}
}
and $h_I(t,z)$, which depends on $t$ and $z^i$, is
a straightforward generalization of the harmonic function associated
with a brane $I$ in a static brane system~\cite{Argurio:1997gt}.

We also assume that the scalar field $\phi$ and
the gauge field strength $F_{(p+2)}$ are given by
\bea
e^{\phi}=\prod_Ih_I^{\epsilon_Ic_I/2},~~~
F_{(p_I+2)}=d(h_I^{-1})\wedge\Omega(\Xsp_I),
  \label{gs:ansatz for fields:Eq}
\ena
where $\Xsp_I$ is the space associated with a brane $I$,
and $\epsilon_I$ is defined by
\Eq{
\epsilon_I=\left\{
\begin{array}{cc}
 + &~{\rm for~the~electric~brane}
\\
 - &~~~{\rm for~the~magnetic~brane}
\end{array} \right. ,
\label{gs:epsilon:Eq}
}
and $
\Omega(\Xsp_I)=dt\wedge dx^{p_1}\wedge \cdots \wedge
dx^{p_I}$  is the volume $(p_I+1)$-form.
The field strength in (\ref{gs:ansatz for fields:Eq}) is written for electric
ansatz, but the final results are basically the same for magnetic ansatz.
In what follows, we write our formulae mainly for electric case with comments
on modifications for magnetic case.

Let us assume~\cite{Argurio:1997gt}
\bea
{\cal A}^{(D-p-3)}\,{\cal B}
 \prod_{\alpha=1}^p\,{\cal C}^{(\alpha)} =1 \,,\nn
{\cal B}^{-1}\prod_{\alpha \in I}\left({\cal C}^{(\alpha)}\right)^{-1}
e^{\epsilon_Ic_I\phi}=h^2_I.
   \label{gs:extremal2:Eq}
\ena
The Einstein equations \Eqref{gs:Einstein equations:Eq} then reduce to
\Eqrsubl{gs:components of Einstein equations:Eq}{
&&
\frac{1}{2}\sum_{I,I'}\left[M_{II'}-2\delta_{II'}
-2\left(a_I-\delta_{II'}a_{I'}\right)
\right]\pd_t\ln h_I\pd_t\ln h_{I'}\nn
&&~~~~+\sum_{I}(2a_I-b_I)h_I^{-1}\pd_t^2h_I
-2\prod_Ih_I^{-1}\sum_{I'}b_{I'}
h_{I'}^{-1}\lap_{\Zsp}h_{I'}=0,
   \label{gs:Einstein equations tt:Eq}\\
&&2\sum_Ih^{-1}_I\pd_t\pd_ih_I+\sum_{I,I'}
\left(M_{II'}-2\delta_{II'}
\right)\pd_t\ln h_I\pd_i\ln h_{I'}=0,
   \label{gs:Einstein equations ti:Eq}\\
&&\prod_{J'}h_{J'}^{-b_{J'}}
\sum_{\gamma}\prod_{J}h_{J}^{c_J^{(\gamma)}}
\sum_{I} \left[c^{(\gamma)}_Ih_I^{-1}\pd_t^2h_I
-\left(c^{(\gamma)}_I\pd_t\ln h_I
-\sum_{I'}c^{(\gamma)}_{I'}\pd_t\ln h_{I'}\right)\pd_t\ln h_I\right]
\nn
&&~~~~-\prod_{J'}h_{J'}^{-a_{J'}}
\sum_{\gamma}\prod_{J}h_{J}^{c_J^{(\gamma)}}\sum_I
c^{(\gamma)}_Ih_I^{-1}\lap_{\Zsp}h_I=0,
   \label{gs:Einstein equations ab:Eq}\\
&&R_{ij}(\Zsp)+\frac{D-2}{2}u_{ij}\prod_Jh_J\sum_{I}
\left[a_I h_I^{-1}\pd_t^2h_I
+ \left\{a_I \pd_t\ln h_I
-\sum_{I'}a_{I'} \pd_t\ln h_{I'}\right\}\pd_t\ln h_I\right]\nn
&&~~~~-\frac{1}{2}u_{ij}\sum_Ih_I^{-1}a_I \lap_{\Zsp}h_I
-\frac{1}{4}\sum_{I,I'}\left(M_{II'}-2\delta_{II'}
\right)\pd_i\ln h_I\pd_j\ln h_{I'}=0 ,
    \label{gs:Einstein equations ij:Eq}
}
where $R_{ij}(\Zsp)$ is the Ricci tensor of the metric $u_{ij}$,
and $M_{II'}$ is given by
\Eqr{
&&M_{II'}\equiv b_Ib_{I'}
+\sum_{\alpha}c^{(\alpha)}_Ic^{(\alpha)}_{I'}
+(D-p-3)a_Ia_{I'} +\frac{1}{2}\epsilon_I\epsilon_{I'}c_Ic_{I'}
\,.
\label{Ein1}
}

Let us consider Eq.~\Eqref{gs:Einstein equations ti:Eq}.
We can rewrite this as
\bea
\sum_{I,I'}\left[ M_{II'}
+2\delta_{II'}\frac{\pa_t\pa_i \ln h_I}{\pa_t \ln h_I \pa_i \ln h_I} \right]
\pa_t \ln h_I \pa_i \ln h_{I'}=0.
\label{Ein2}
\ena
In order to satisfy this equation for arbitrary coordinate values and
 independent
functions $h_I$, the second term in the square bracket must be constant:
\bea
\frac{\pa_t\pa_i \ln h_I}{\pa_t \ln h_I \pa_i \ln h_I}=k_I
\,.
\label{Ein3}
\ena
Then in order for \Eqref{Ein2} to be satisfied identically, we must have
\bea
M_{II'}+2k_I \delta_{II'}=0.
\label{const1}
\ena
Using Eqs.~\Eqref{gs:parameter c:Eq}, \Eqref{gs:paremeter:Eq} and \Eqref{Ein1},
we get
\bea
M_{II} &=&
 (p_I+1)b_I^2 + (p-p_I)a_I^2
+(D-p-3)a_I^2 +\frac12 c_I^2 \nonumber \\
&=& 2.
\ena
This means that the constant $k_I$ in Eq.~\Eqref{const1} is $k_I=-1$, namely
\bea
M_{II'}= 2\delta_{II'}.
\label{diag}
\ena
It then follows from Eq.~\Eqref{Ein3} that
\bea
\pa_i \pa_t[h_I(t,z)]=0
\,.
\ena
As a result, the warp factor $h_I$ must be separable as
\Eq{
h_I(t, z)= K_I(t)+H_I(z)
\,.
  \label{gs:form of warp factor:Eq}
}

For $I\neq I'$, Eq.~\Eqref{diag} gives the intersection rule
on the dimension $\bar{p}$ of the intersection for each pair of branes
$I$ and $I'$ $(\bar{p}\leq p_I, p_{I'})$~\cite{Gauntlett:1996pb,Tseytlin:1996hi,
Argurio:1997gt,Ohta:1997gw}:
\Eq{
\bar{p}=\frac{(p_I+1)(p_{I'}+1)}{D-2}-1-
\frac{1}{2}\epsilon_Ic_I\epsilon_{I'}c_{I'}.
\label{isr}
}

Let us next consider the gauge field.
Under the ansatz~\Eqref{gs:ansatz for fields:Eq} for electric background,
we find
\Eq{
dF_{(p_I+2)}=h_I^{-1}(2\pd_i \ln h_I \pd_j \ln h_I
+ h_I^{-1}\pd_i\pd_j h_I)
dz^i\wedge dz^j\wedge\Omega(\Xsp_I)=0.
}
Thus, the Bianchi identity is automatically satisfied.
Also the equation of motion for the gauge field becomes
\Eqr{
d\left[e^{-c_I\phi}\ast F_{(p_I+2)} \right]
&=&-d\left[\pd_ih_I \left\{\ast_{\Zsp}dy^i\wedge
\ast_{\Xsp}\Omega(\Xsp_I)\right\}\right]\nn
&=&-\left(\pd_t\pd_ih_Idt+\lap_{\Zsp}h_Idy^i\right)
\wedge\left[\ast_{\Zsp}dy^i\wedge
\ast_{\Xsp}\Omega(\Xsp_I)\right]=0,
 }
where $\ast_{\Xsp}$, $\ast_{\Zsp}$ denotes the Hodge dual
operator on $\Xsp(\equiv \cup_{I}X_I)$ and $\Zsp$, respectively,
and we have used Eqs.~\Eqref{gs:extremal2:Eq}.
Hence we again find the condition~\Eqref{gs:form of warp factor:Eq} and
\bea
\label{gs:solution of gauge field equations:Eq}
\lap_{\Zsp}h_I&=&0.
\ena
We note that the roles of the Bianchi identity and field equations are
interchanged for magnetic ansatz~\cite{Argurio:1997gt,Ohta:1997gw},
but the net result is the same.

Let us finally consider the scalar field equation.
Substituting the scalar field and the gauge field in \Eqref{gs:ansatz for fields:Eq},
and the warp factor \Eqref{gs:form of warp factor:Eq}
into the equation of motion for the scalar field
\Eqref{gs:scalar field equation:Eq}, we obtain
\Eqr{
&&-\prod_{I''}h_{I''}^{-b_{I''}}
\sum_{I}\epsilon_Ic_I\left[h_I^{-1}\pd_t^2K_I
+\pd_t\ln h_I\sum_{I'}\pd_t\ln h_{I'}-
(\pd_t\ln h_I)^2
\right]\nn
&&~~~~~
+\prod_{I''}h_{I''}^{-a_{I''}}\sum_Ih_I^{-1}
\epsilon_Ic_I\lap_{\Zsp} H_I=0.
  \label{gs:scalar field equation2:Eq}
}
This equation is satisfied if
\Eqrsubl{enough}{
&&\pd_t^2K_I=0,
\label{eq_KI}
\\
&&\triangle_{\Zsp}H_I=0,
   \label{5}\\
&&\sum_{I}\epsilon_Ic_I\left[
\pd_t\ln h_I\sum_{I'}
\pd_t\ln h_{I'}-(\pd_t\ln h_I)^2
\right]=0.
\label{gs:condition for the scalar field:Eq}
}
Eq. (\ref{eq_KI}) gives
$K_I=A_I t+B_I$, where where $A_I$ and $B_I$ are integration constants.
Eq.~\Eqref{gs:condition for the scalar field:Eq}
can be satisfied {\it only if there is only one function $h_I$ depending
on both $z^i$ and $t$, which we denote with the subscript $\tilde I$,
and other functions are either dependent on $z^i$ or constant.}
Hence we have
\Eqr{
&&K_{\tilde I}=A_{\tilde I}\, t+B_{\tilde I},
\nn
&&K_{I}=B_{I},~~~(I\neq \tilde I).
\label{sol_KI}
}

The remaining
Einstein equations~\Eqref{gs:components of Einstein equations:Eq} now reduce to
\Eqrsubl{red}{
&&\sum_{I,I'} \left[a_I- \delta_{II'}a_{I'}\right]
\pd_t\ln h_I\pd_t\ln h_{I'}=0,
\label{1}\\
&&\sum_{I}\left[
\pd_t\ln h_I\sum_{I'}
\pd_t\ln h_{I'}-(\pd_t\ln h_I)^2
\right]
=0,
\label{2}\\
&&R_{ij}(\Zsp)=0.
\label{3}
}
Obviously the first two sets of equations~\Eqref{1} and \Eqref{2} are
automatically satisfied by our solutions in which there is only one
function $h_{\tilde I}$ depending on both $t$ and $z^i$.
Given the set of solutions to Eqs.~\Eqref{gs:form of warp factor:Eq},
\Eqref{5}, \Eqref{sol_KI}, and \Eqref{3},
we have thus obtained general intersecting
dynamical brane solutions~\Eqref{gs:metric:Eq}.
For static (time-independent) case, our solutions are consistent with
the harmonic function rule~\cite{Tseytlin:1996bh}, but are more general
with spacetime-dependent functions.
Note that the internal space is not warped~\cite{Kodama:2005cz}
if the function $H_I$ is trivial.

As a special example, we consider the case
\Eq{
\quad u_{ij}=\delta_{ij}\,,
 \label{gs:special metric of solution:Eq}
}
where $\delta_{ij}$ is the $(D-p-1)$-dimensional Euclidean metric.
In this case, the solution for $H_I$ can be obtained explicitly as
\Eqr{
H_I(z)=1 +\sum_{k}\frac{Q_{I, \,k}}{|\bm{z}- \bm{z}_{k}|^{D-p-3}},
}
where  $Q_{I,\, k}$'s are constant parameters and $\bm{z}_{k}$
represent the positions of the branes in Z space.\footnote{
Here we show the solution without compactification of $\Zsp$ space.
One may also discuss the case that $q$-dimensions of $\Zsp$ space are
smeared, which gives the different power of harmonics, i.e.
${|\bm{z}- \bm{z}_{k}|^{-(D-p-3-q)}}$ ($q\leq D-p-2$). }
For $K_{\tilde I}=0 ~(A_{\tilde I}=B_{\tilde I}=0$),
the metric describes the known static and extremal multi-black hole
solution with black hole charges
$Q_{I,\, k}$~\cite{Argurio:1997gt,Ohta:1997gw,Argurio:1998cp}.

\section{Classification of dynamical intersecting M-branes}
\label{sec:classification}

Now, we give a classification of multiple intersections of
M-branes in eleven dimensions. The intersections of D-branes and other branes
can be obtained by dimensional reductions and T-duality.
We look for the possible configurations of intersecting branes
by use of \Eqref{isr}.
It turns out that no configuration is possible for more than
eight branes~\cite{Bergshoeff:1996rn}.
In what follows, we present explicit solutions.
The case with M-waves or KK-monopoles
will be discussed later (\S~\ref{Mwave_KKmonopole}).

\subsection{Dynamical intersecting M-branes}

In our solutions~\Eqref{gs:metric:Eq}, only one time-dependent brane is
allowed, which we denote by $\tilde I$.
Then we have
\bea
h_{\tilde I}=h_{\tilde I}(t, z)\equiv A_{\tilde I} \, t+H_{\tilde I}(z)
\,.
\ena
Here we set $B_{\tilde I}=0$ without loss of generality.
We write the solution as
\bea
ds^2 = {\cal A}(t,z)
\left[-g_0(t,z)dt^2+
\sum_{\tilde \alpha \in \tilde I}
g_{\tilde \alpha}(t,z)(dx^{\tilde \alpha})^2
+\sum_{\alpha \in \hspace{-.4em}/ \tilde I}
g_{\alpha}(z)(dy^{\alpha})^2+u_{ij}(z)dz^idz^j
\right],\nn
\label{metric_Mbrane}
\ena
with
\bea
{\cal A}&=&\left[A_{\tilde I}t+H_{\tilde I}(z)\right]^{a_{\tilde I}}
\prod_{I\neq \tilde I}H_I(z)^{a_{I}}~~,~~~
g_0\,=\,\left[A_{\tilde I}t+H_{\tilde I}(z)\right]^{-1}
\prod_{I\neq \tilde I}H_I(z)^{-1}, \nn
g_{\tilde \alpha}&=&
\left[A_{\tilde I}t+H_{\tilde I}(z)\right]^{-1}
\prod_{I\neq \tilde I}
H_I^{-\gamma_I^{(\tilde \alpha)}}~~,~~~
g_{\alpha}\, =\,
\prod_{I\neq \tilde I}H_I^{-\gamma_I^{(\alpha)}}
\,,
\ena
where
\bea
a_{\tilde I}={p_{\tilde I}+1\over D-2}~~\,,
~~{\rm and}~~~
\gamma^{(\alpha)}_I=\left\{
\begin{array}{cc}
1&~{\rm for}~~\alpha\in I\\ 0 &~{\rm for}~~\alpha
\in \hspace{-.8em}/ I
\end{array} \right. .
\ena
Here we divide the coordinates of brane world-volume ($\{x^\alpha\}$)
into two parts ($\{x^{\tilde \alpha}\}, \{y^\alpha\}$):
the first are the $p_{\tilde I}$-dimensional coordinates $x^{\tilde \alpha}$
which describe the time-dependent brane $\tilde I$, and
the second are the ($p-p_{\tilde I}$)-dimensional coordinates $y^\alpha$
which represent the remaining space of the brane world-volume.

Let us now give one simple example of M5-M5 brane system.
The intersection rule~(\ref{isr}) gives the following brane configuration:
\begin{table}[ht]
\caption{M5-M5 brane system}
{\scriptsize
\begin{center}
\begin{tabular}{|c|c|c|c|c|c|c|c|c|c|c|c||c|}
\hline
&0&1&2&3&4&5&6&7&8&9&10&$\tilde{I}$
\\
\hline
M5 & $\circ$ & $\circ$ & $\circ$ & $\circ$ & $\circ$ & $\circ$ &&&&&& $\surd$
\\
\cline{2-13}
M5 & $\circ$ & $\circ$ & $\circ$ & $\circ$ &&& $\circ$ & $\circ$ & & &&
\\
\cline{1-13}
& $t$ & $x^{\tilde 1}$ & $x^{\tilde 2}$ & $x^{\tilde 3}$ &$x^{\tilde 4}$
&$x^{\tilde 5}$& $y^6$ & $y^7$ &$z^1$ &$z^2$ &$z^3$&
\\
\hline
\end{tabular}
\end{center}
}
\end{table}

The mark $\surd$ in the table shows which brane is time dependent, though in this case
there is no difference whichever of the two M5's is chosen.
The metric is then given by
\bea
ds^2&=& (h_{\tilde 5}H_5)^{2/3}\left[(h_{\tilde 5}H_5)^{-1}
\left(-dt^2+\sum_{\tilde \alpha=1}^3(dx^{\tilde \alpha})^2\right)
     +h_{\tilde 5}^{-1} \sum_{\tilde \alpha=4}^5(dx^{\tilde \alpha})^2
\right.\nn
     & &
\hskip 2cm
\left.+H_5^{-1} \sum_{\alpha=6}^7(dy^\alpha)^2
     +u_{ij}dz^idz^j\right]
\,,
\ena
that is
\begin{eqnarray}
&&{\cal A}=(h_{\tilde 5}H_5)^{2/3}\,, \nn
&& g_0=g_{\tilde 1}=g_{\tilde 2}=g_{\tilde 3}=(h_{\tilde 5}H_5)^{-1}
~\,,~~
g_{\tilde 4}=g_{\tilde 5}=h_{\tilde 5}^{-1},
\\
&&g_{1}=g_{2}=H_5^{-1}, \nonumber
\end{eqnarray}
where
\begin{eqnarray}
h_{\tilde 5}=A_{\tilde 5}t+H_{\tilde 5}(z)
\,.
\end{eqnarray}
The form field is given by
\begin{eqnarray}
F_{(4)}&=-&\ast_{\Zsp} \left(dh_{\tilde 5}\right)\wedge dx^{\tilde 4}\wedge
dx^{\tilde 5}-\ast_{\Zsp}\left(dH_5\right)\wedge dy^6\wedge dy^7\,,
\end{eqnarray}
where $\ast_{\Zsp}$ is the Hodge dual operator
in the three-dimensional ${\Zsp}$ space.

The complete classification and explicit metrics for intersecting brane systems
are summarized in Appendix \ref{Appendix:table_intersectingMbranes}.

\subsection{Spacetime structure of the intersecting branes}

Near branes ($|\bm z|\sim 0$), the spacetime structure is the same as
that of the static solution unless the dimension of $\Zsp$ space is one.
This is because the metric components diverge as $|\bm z|\rightarrow 0$
and the static harmonic parts dominate the time-dependent terms.
In that case, we know that
M2-M2, M2-M5, M2-M2-M2, M5-M5-M5, M2-M2-M5-M5
systems are regular on the branes.

If $\Zsp$ space is one-dimensional, then we have
$h_{\tilde I}=A_{\tilde I}t+\sum_k Q_{I,\,k}|z- z_k|$.
Hence any points on the branes ($z=z_k$) are regular,
and time dependent.

Even if the near-brane structure is regular, we expect
another type of singularity may appear at $h_{\tilde I}(t,z)=0$.
Since  $h_{\tilde I}$ is a linear function of $t$, it vanishes
once for any position $\bm z$ at $t=-H_{\tilde I}(z)/A_{\tilde I}$.

When we take the limit of   $H_{\tilde I}\rightarrow 0$ (or finite)
as $|\bm z|\rightarrow \infty$ for ${\rm dim}(\Zsp)>1$ (or
$|z|$ is finite for ${\rm dim}(\Zsp)=1$),
the spacetime turns out to be time dependent and homogeneous.
To see its dynamical behaviour,
we introduce a new time coordinate
\bea
\tau= \tau_0 (A_{\tilde I} t)^{(a_{\tilde I}+1)/2},
\label{newtime}
\ena
where $\tau_0=\frac{2}{A_{\tilde I}(a_{\tilde I}+1)}$.
The asymptotic solution is rewritten as
\bea
ds^2=
-d\tau^2+\left({\tau\over \tau_0}\right)^{2q_{\tilde I}}
\sum_{\tilde \alpha}(dx^{\tilde \alpha})^2
+\left({\tau\over \tau_0}\right)^{2q_{\tilde I  \hspace{-.3em}\setminus}}
\left(\sum_{\alpha}(dy^{\alpha})^2+u_{ij}dz^idz^j
\right)
\,,
\label{gs:class:Eq}
\ena
where
\bea
q_{\tilde I}={a_{\tilde I}-1\over a_{\tilde I}+1}
=-{D-p_{\tilde I}-3\over D+p_{\tilde I}-1}
\,,~~
q_{\tilde I \hspace{-.4em}\setminus}={a_{\tilde I}\over a_{\tilde I}+1}
={p_{\tilde I}+1\over D+p_{\tilde I}-1}.
\ena
More explicitly, for the case of M-theory ($D=11$), we find
\Eqrsubl{exp}{
&&
a_{\tilde I}=1/3, \,
q_{\tilde I}=-1/2, \,
q_{\tilde I \hspace{-.4em}\setminus}=1/4,
~~{\rm for}  ~\tilde I={\rm M2} ~({p_{\tilde I}}=2)
\,,
\\
&&
a_{\tilde I}=2/3,\,
q_{\tilde I}=-1/5,\,
q_{\tilde I \hspace{-.4em}\setminus}=2/5,
~~{\rm  for}  ~\tilde I={\rm M5}~({p_{\tilde I}}=5)
\,.
}
Hence, we find a Kasner-like expansion:
\Eqrsubl{kl}{
&&
p_{\tilde I}\,q_{\tilde I}+p_{{\tilde I \hspace{-.4em}\setminus}}\,
q_{{\tilde I\hspace{-.4em}\setminus}}=1\,,
\label{Kas1}
\\
&&
p_{\tilde I}(q_{\tilde I})^2+p_{{\tilde I \hspace{-.4em}\setminus}}
(q_{{\tilde I \hspace{-.4em}\setminus}})^2=1
\,.
\label{Kas2}
}
where $p_{{\tilde I \hspace{-.4em}\setminus}}=(D-p_{\tilde I}-1)$ is the dimension of
the space volume  perpendicular to the $\tilde I$-brane world-volume.
Eq. (\ref{Kas1}) is always satisfied for any brane configuration,
 but Eq. (\ref{Kas2}) is true only for
M-theory because no dilaton appears.

This time dependence is also correct if we fix the position in $\Zsp$ space,
although the metric is locally inhomogeneous in the bulk space.

\section{Applications to cosmology and black holes}
\label{application}

\subsection{Cosmology}
\label{application_cosmology}

Now we discuss how these solutions are applied to our physical world.
Since we consider time-dependent solutions, it is natural to discuss cosmology.
Suppose that our three-dimensional universe is a part of branes.
Since our universe is isotropic and homogeneous,
same branes must contain this whole three dimensions.
Hence we should look for whether there is a solution with
an isotropic and homogeneous three space from a list of our solutions given
in Appendix \ref{Appendix:table_intersectingMbranes}.
Note that this does not mean that the three space must be contained in all branes.
We find just six cases, i.e.,
M2-M5, M5-M5, M5-M5-M5, M2-M5-M5, M2-M2-M5, and M2-M2-M5-M5 brane systems.
In some cases, we have two different expansion laws for our universe
depending on whether the brane on which our world exists is time dependent or not.

We then compactify some dimensions to fit our three space.
We assume that our universe is one of the branes
(or its three-dimensional part), which can be
the time-dependent one ($\tilde I$) or
the static one  ($I(\neq \tilde I)$).
Hence our universe stays at a constant position in the bulk space
($\bm{z}=\bm{z}_k$).
Note that among the above spacetimes,
only M2-M5 and M2-M2-M5-M5 brane systems are regular on the branes.
For other configurations, the curvature diverges there. Hence one need invoke
a mechanism to avoid singularity if our world is confined on the brane.

We describe our three space  $\Xi$ by the coordinates
$\bm{\xi}=(\xi^1, \xi^2, \xi^3)$.
There are two possibilities: One is that $\Xi$ belongs to
some part of the time-dependent brane world-volume $\Xsp_{\tilde I}$
(case 1), and the other is that $\Xi$ is contained in a part of
only static brane world-volume $\Ysp_{I}(I\neq {\tilde I})$,
which does not belong to  $\Xsp_{\tilde I}$ (case 2).

For the case 1, the metric (\ref{metric_Mbrane}) is described by
\bea
ds^2=ds_4^2+ds_{p-3}^2+ds_{\rm bulk}^2
\,,
\ena
where
\bea
ds_4^2&=&{\cal A}\left[
-g_0dt^2+g_\xi\sum_{\tilde \alpha \in \Xi}(dx^{\tilde\alpha})^2 \right],
\nn
ds_{p-3}^2&=&
{\cal A}\left[
\sum_{\tilde \alpha \in \hspace{-.4em}/
 \Xi,\, \tilde \alpha \in \tilde I}
g_{\tilde \alpha}(dx^{\tilde \alpha})^2
+\sum_{\alpha \in \hspace{-.4em}/ \tilde I}g_\alpha
(dy^\alpha)^2 \right],
\nn
ds_{\rm bulk}^2&=&{\cal A}\, u_{ij}dz^idz^j
\,.
\ena
{}From our ansatz, $g_{\tilde \alpha}$'s for our three space
(${\tilde \alpha}\in \Xi$) are the same, which we denote $g_\xi$.
$ds_{p-3}^2$ is the part of compactified brane world-volume,
and $ds_{\rm bulk}^2$ describes the empty bulk space.

We have to describe our 4-dimensional universe in the Einstein frame,
which is given by
\bea
d\bar{s}_{4}^2
&\equiv&
\prod_{\tilde \alpha \in \hspace{-.4em}/
 \Xi,\, \tilde \alpha \in \tilde I}
\left({\cal A}g_{\tilde \alpha}\right)^{1/2}
\prod_{\alpha \in \hspace{-.4em}/  \tilde I}
\left({\cal A}g_{\alpha}\right)^{1/2}ds_4^2
\nn
&=&
h_{\tilde I}^{~s_{\tilde I}}(t,z)
 F_{\tilde I}(z)\left[-f_0(z)dt^2+f_\xi(z) d\bm{\xi}^2
\right]
\,,
\ena
where
\bea
s_{\tilde I}
&=&{1\over 2}\left[-(p_{\tilde I}-1)+{(p-1)\over (D-2)}(p_{\tilde I}+1)\right],
\nn
F_{\tilde I}&=&
\prod_{I\neq \tilde I}H_I^{(p_{I}+1)(p-1)\over 2(D-2)}
\times
\prod_{\tilde \alpha \in \hspace{-.4em}/
 \Xi,\, \tilde \alpha \in \tilde I}
\left(\prod_{I\neq \tilde I} H_I^{-\gamma_I^{(\tilde \alpha)}/2}\right)
\times
\prod_{\alpha \in \hspace{-.4em}/ \tilde I}
\left(\prod_{I\neq \tilde I} H_I^{-\gamma_I^{(\alpha)}/2}\right),
\nn[.5em]
f_0&=&\prod_{I\neq \tilde I} H_I^{-1}
~~,~~~
f_\xi\, =\, \prod_{I\neq \tilde I} H_I^{-\gamma_I^{(\xi)}} .
\ena
Here note that the middle factor in $F_{\tilde I}$ has the exponent
$\gamma_I^{(\tilde \alpha)}$ which is nonvanishing for the case where
the coordinate $x^{\tilde\a}$ belongs to time-dependent brane as well as
time-independent $I$ brane.

For the case 2, we have
\bea
ds_4^2&=&{\cal A}\left[
-g_0dt^2+g_\xi\sum_{\alpha \in \Xi, \in \hspace{-.4em}/ \tilde I}
(dy^{\alpha})^2 \right],
\nn
ds_{p-3}^2&=&
{\cal A}\left[
\sum_{\tilde \alpha \in \tilde I}
g_{\tilde I}(dx^{\tilde \alpha})^2
+\sum_{\alpha \in \hspace{-.4em}/ \Xi,
 \tilde I}g_\alpha
(dy^\alpha)^2 \right],
\nn
ds_{\rm bulk}^2&=&{\cal A}u_{ij}dz^idz^j
\,.
\ena
Hence the 4-dimensional metric of our universe in the Einstein frame is
\bea
d\bar{s}_{4}^2
&\equiv&
\prod_{\tilde \alpha \in \tilde I}
\left({\cal A}g_{\tilde \alpha}\right)^{1/2}
\prod_{\alpha \in \hspace{-.4em}/ \Xi,
 \in \hspace{-.4em}/  \tilde I}
\left({\cal A}g_{\alpha}\right)^{1/2}ds_4^2
\nn
&=&
h_{\tilde I}^{~s_{\tilde{I}\hspace{-.3em}\setminus}}(t,z)
 F_{\tilde{I}\hspace{-.4em}\setminus}(z)
\left[-f_0(z)dt^2+h_{\tilde I}(t,z)f_\xi(z) d\bm{\xi}^2
\right]
\,,
\ena
where
\bea
s_{\tilde{I}\hspace{-.4em}\setminus}
&=&{1\over 2}\left[-(p_{\tilde I}+2)+{(p-1)\over (D-2)}(p_{\tilde I}+1)\right],
\nn
F_{\tilde{I}\hspace{-.4em}\setminus}&=&
\prod_{I\neq \tilde I}H_I^{(p_{\tilde I}+1)(p-1)\over 2(D-2)}
\times
\prod_{\tilde \alpha \in \tilde I}
\left(\prod_{I\neq \tilde I} H_I^{-\gamma_I^{(\tilde \alpha)}/2}\right)
\times
\prod_{\alpha \in \hspace{-.4em}/ \Xi,
 \in \hspace{-.4em}/  \tilde I}
\left(\prod_{I\neq \tilde I} H_I^{-\gamma_I^{(\alpha)}/2}
\right).
\ena

Since we fix our universe at some position in the bulk $\Zsp$ space,
$\bm{z}$ is constant in the above metric.
Hence we find the isotropic and homogeneous universe.
We introduce the cosmic time $\tau$, which is defined by
\bea
{\tau} =
\left\{
\begin{array}{cc}
\tau_{\tilde I}(A_{\tilde I}t)^{(s_{\tilde I} +2)/ 2}&
\hspace{.5cm} {\rm for ~the~ case~ 1} \\[1em]
\tau_{\tilde I\hspace{-.4em}\setminus}(A_{\tilde I}t)^{(s_{\tilde I
\hspace{-.3em}\setminus} +2)/ 2}&\hspace{.5cm}
  {\rm for ~the ~case~ 2}
\end{array}
\right.
\,,
\ena
where $\tau_{\tilde I}=2/[A_{\tilde I}(s_{\tilde I} +2)]$ and
$\tau_{\tilde I\hspace{-.4em}\setminus}
=2/[A_{\tilde I}(s_{\tilde I\hspace{-.4em}\setminus} +2)]$,
respectively.
The scale factor of the universe is given by
\bea
a_{\tilde I}&=&(A_{\tilde I}t) ^{s_{\tilde I}/2}=
\left({\tau\over \tau_0}\right)^{\beta_{\tilde I}},
\nn
a_{\tilde I\hspace{-.4em}\setminus}&=&
(A_{\tilde I}t)^{(s_{\tilde I\hspace{-.3em}\setminus}+1)/2}=
\left({\tau\over \tau_0}\right)^{\beta_{\tilde I\hspace{-.3em}\setminus}}
\,,
\ena
where
\bea
\beta_{\tilde I}={{s_{\tilde I}\over (s_{\tilde I}+2)}}
~~,~{\rm and}~~~
\beta_{\tilde I\hspace{-.35em}\setminus}={
{(s_{\tilde I\hspace{-.35em}\setminus}+1)
\over (s_{\tilde I\hspace{-.35em}\setminus}+2)}}.
\label{power_exponent}
\ena

The power of the cosmological solution for each possible model
is listed in Table~\ref{table_power}.
Since the time dependence in the metric comes from only one
M-brane (or D-brane) in the intersections, the obtained expansion law
may be too simple.
In fact, we find the Minkowski space, which is static, in almost
every case.

\begin{table}
\caption{
The power exponent $\beta_{\tilde I}$ ( or
$\beta_{\tilde I\hspace{-.4em}\setminus}$ ) of the scale factor
$a_{\tilde I}$ ( or $a_{\tilde I\hspace{-.4em}\setminus}$ ) of
possible $4$-dimensional cosmological model is given, i.e. $a
\propto \tau^\beta$, where $\tau$ is the cosmic time.
The last three columns are for the case of smeared and compactified bulk space.
}
\begin{center}
{\scriptsize
\begin{tabular}{|c||c||c|c|c||c|c|c|}
\hline
&branes&${\rm dim}(\Zsp)$&$s_{\tilde I}$ or
$s_{\tilde I\hspace{-.4em}\setminus}$
&$\beta_{\tilde I}$ or
$\beta_{\tilde I\hspace{-.4em}\setminus}$&$\beta_{\tilde I}^{(1)}$ or
$\beta_{\tilde I\hspace{-.4em}\setminus}^{(1)}$ &$\beta_{\tilde I}^{(2)}$ or
$\beta_{\tilde I\hspace{-.4em}\setminus}^{(2)}$ &$\beta_{\tilde I}^{(3)}$ or
$\beta_{\tilde I\hspace{-.4em}\setminus}^{(3)}$
\\
\hline
\hline
&M2-M5&4&$-1/3$&$-1/5$&0&1/7&1/4
\\
\cline{2-8}
&M5-M5&3&0&0&1/7&1/4&$-$
\\
\cline{2-8}
case 1
&M5-M5-M5&1&2/3&1/4&$-$&$-$&$-$
\\
\cline{2-8}
(${\tilde I}=$M5)
&M2-M5-M5&3&0&0&1/7&1/4&$-$
\\
\cline{2-8}
&M2-M2-M5&3&0&0&1/7&1/4&$-$
\\
\cline{2-8}
&M2-M2-M5-M5&3&0&0&1/7&1/4&$-$
\\
\hline
\hline
&M2-M5&4&$-7/6$&$-1/5$&0&1/7&1/4
\\
\cline{2-8}
case 2&M2-M5-M5&3&$-1$&0&1/7&1/4&$-$
\\
\cline{2-8}
(${\tilde I}=$M2)
&M2-M2-M5&3&$-1$&0&1/7&1/4&$-$
\\
\cline{2-8}
&M2-M2-M5-M5&3&$-1$&0&1/7&1/4&$-$
\\
\hline
\end{tabular}
}
\label{table_power}
\end{center}
\end{table}

In order to find an expanding universe, one may have to smear and
compactify the vacuum bulk space as well as the brane world-volume.
Suppose $k$-dimensions of the bulk $\Zsp$ space are smeared and compactified,
where $k<{\rm dim}(\Zsp)=D-p-1$.
The metric in the Einstein frame is multiplied by
the extra factor ${\cal A}^{k/2}$.
As a result, we find new exponents of the metric are
\bea
s_{\tilde I}^{(k)}&=& s_{\tilde I}+{k(p_{\tilde I}+1)\over 2(D-2)},
\nn
s_{\tilde I\hspace{-.4em}\setminus}^{(k)}&=&
 s_{\tilde I\hspace{-.4em}\setminus}
+{k(p_{\tilde I}+1)\over 2(D-2)}.
\ena
The power of the scale factor is given by the same equations
(\ref{power_exponent}) by replacing $ s_{\tilde I}$ with $s_{\tilde I}^{(k)}$
(or $ s_{\tilde I\hspace{-.4em}\setminus}$
with $s_{\tilde I\hspace{-.4em}\setminus}^{(k)}$).
We also show these explicit powers in Table~\ref{table_power}.
However, even for the fastest expanding case $a\propto \tau^{1/4}$,
the power is too small to give a realistic expansion law such as that in
the matter dominated era ($a\propto \tau^{2/3}$)
or that in the radiation dominated era ($a\propto \tau^{1/2}$).

Hence we conclude that
in order to find a realistic expansion of the universe
in this type of models,
one have to include additional ``matter" fields on the brane.

\subsection{Time-dependent black holes}
\label{application_BH}

Since the static (or stationary) intersecting brane system
describes the microstate of a black hole,
it may be natural to apply the present solutions
to a time-dependent spacetime with a black hole.
In this case, just as the case of a static black hole, we
should compactify all brane world-volume, and
obtain the $d$-dimensional spacetime, where $d\equiv D-p={\rm dim}(\Zsp)+1$.
Our metric is described as
\bea
ds^2=ds_d^2
+ds_p^2
\,,
\ena
where
\bea
ds_d^2&=&{\cal A}\left[-g_0 dt^2+u_{ij}dz^idz^j\right],
\nn
ds_p^2&=&{\cal A}\left[
\sum_{\tilde\alpha \in\tilde I}g_{\tilde \alpha}(dx^{\tilde\alpha})^2
+\sum_{\alpha \in\hspace{-.4em}/
\tilde I}g_{\alpha}(dy^{\alpha})^2
\right]
\,.
\ena
The compactification of  $ds_p^2$ gives the effective $d$-dimensional
spacetime, whose metric in the Einstein frame is given by
\bea
d\bar s_{d}^2=\prod_{{\tilde\alpha \in\tilde I}}
\left({\cal A}g_{\tilde \alpha}\right)^{1/(d-2)}
\prod_{\alpha \in\hspace{-.4em}/ \tilde I}
\left({\cal A}g_{\alpha}
\right)^{1/(d-2)}\,{\cal A}
\left(-g_0 dt^2+u_{ij}dz^idz^j\right)
\,,
\ena
which is rewritten explicitly as
\bea
d\bar s_{d}^2=h_{\tilde I}^{s_{\rm BH}} F_{\rm BH}(z)\left(
-f_0(z)dt^2+h_{\tilde I}(t,z)u_{ij}dz^idz^j\right)
\,,
\label{metric_BH}
\ena
where
\bea
{s_{\rm BH}}&=&-{d-3\over d-2},
\nn
F_{\rm BH}(z)&=&\prod_{I\neq \tilde I}
H_I^{{(p_{I}+1)(p+2)\over (d-2)(D-2)}}
\prod_{\tilde \alpha\in \tilde I}\left(\prod_{I\neq \tilde I}
H_I^{-\gamma_I^{(\tilde \alpha)}/(d-2)}\right)
\prod_{\alpha\in \hspace{-.4em}/ \tilde I}\left(
\prod_{I\neq \tilde I}
H_I^{-\gamma_I^{(\alpha)}/(d-2)}\right)
\,.
\ena

We look for a four or higher dimensional ``black hole",
i.e. $d\geq 4$, or equivalently
${\rm dim}(\Zsp)\equiv D-p-1\geq 3$. In M-theory, this implies that $p\leq 7$.
The corresponding brane systems are
M2-M2, M2-M5, M5-M5, M5-M5-M5, M2-M5-M5, and M2-M2-M5-M5.

The near-brane geometry is the same as the static one because
$h_{\tilde I}\rightarrow H_{\tilde I}(z)$  as $\bm{z}\rightarrow \bm{z}_k$
and then the geometry approaches the static solution.
If it has a horizon geometry, we can regard
the present time-dependent solution as a black hole.
We know that only two cases (M2-M2-M2, M2-M2-M5-M5)
give regular black hole spacetimes in the static limit.

On the other hand, the asymptotic structure is completely different.
The static solution has an asymptotically flat geometry, but the present
solution is time dependent.
In fact, setting $h_{\tilde I}=t/t_0+H_{\tilde I}$,
from Eq. (\ref{metric_BH})
in the limit of $|\bm{z}|  \rightarrow \infty$,
we find
\bea
d\bar s_{d}^2&=& \left({t\over t_0}\right)^{s_{\rm BH}}
\left[
-dt^2+\left({t\over t_0}\right)\,u_{ij}dz^idz^j\right]
\nonumber
\\[1em]
&=&-d\tau^2+a_{\rm BH}^2(\tau)\, u_{ij}dz^idz^j
\,,
\ena
where
\bea
a_{\rm BH}= \left({\tau\over \tau_0}\right)^{\beta_{\rm BH}},
\ena
with
\bea
\beta_{\rm BH}={s_{\rm BH}+1\over s_{\rm BH}+2}={1\over d-1}
~~,~~~
\tau_0={2\over s_{\rm BH}+2}\,t_0={2(d-2)\over d-1}t_0
\,.
\ena
Hence our solution approaches  asymptotically
 the FRW universe
with the scale factor $a_{\rm BH}$.
So, if the static solution gives a black hole, then we can regard
the present solution as a black hole in the expanding universe.
In Table \ref{table_BH}, we show a list of the power exponent
of asymptotic expanding universe for the possible black hole
(or black object) model.

\begin{table}
\caption{The power exponent of the asymptotic expansion for
``BH" spacetime. Only the brane systems marked in the column ``BH"
have regular horizons.}
\begin{center}
{\scriptsize
\begin{tabular}{|c|c||c||c|c||c||c|}
\hline
branes&$d$&$\tilde I$&$s_{\rm BH}$
&$\beta_{\rm BH}$&
$\beta_{\rm BH}^{(k)}$ &BH
\\
\hline
\hline
M2-M2&7&M2&$-4/5$&1/6&1/5, 1/4, 1/3 ($k=1, 2, 3$)&
\\
\hline
M2-M5&5&M2&$-2/3$&1/4&1/3 ($k=1$)&
\\
\cline{3-6}
&&M5&$-2/3$&1/4&1/3 ($k=1$)&
\\
\hline
M5-M5&4&M5&$-1/2$&1/3&$-$&
\\
\hline
M5-M5-M5&4&M5&$-1/2$&1/3&$-$&
\\
\hline
M2-M5-M5&4&M2&$-1/2$&1/3&$-$&
\\
\cline{3-6}
& &M5&$-1/2$&1/3&$-$&
\\
\hline
M2-M2-M5&4&M2&$-1/2$&1/3&$-$&
\\
\cline{3-6}
& &M5&$-1/2$&1/3&$-$&
\\
\hline
M2-M2-M2&5&M2&$-2/3$&1/4&1/3 ($k=1$)&$\surd$
\\
\hline
M2-M2-M5-M5&4&M2&$-1/2$&1/3&$-$&$\surd$
\\
\cline{3-6}
& &M5&$-1/2$&1/3&$-$&$\surd$
\\
\hline
\end{tabular}
}
\label{table_BH}
\end{center}
\end{table}

If we smear and compactify the vacuum bulk $\Zsp$ space just as the case of cosmology,
we find the different power exponent of the scale factor,
which is also shown in Table \ref{table_BH}.
As a result, we always find the same power $\beta_{\rm BH}=1/(d-1)$ for
a $d$-dimensional black hole (or black object).
This power exponent is obtained for the universe filled by
stiff matter whose equation of state is $P=\rho$.
Therefore we may regard the present $d$-dimensional solution as
a time-dependent black hole in the stiff-matter dominated universe.

Here we give one explicit example of M2-M2-M5-M5 brane system.
We assume that one M2 brane is time-dependent.
\bea
d\bar s_{4}^2&=&-(h_{\tilde 2}H_2H_5 H_{5'})^{-1/2} dt^2
+(h_{\tilde 2}H_2H_5 H_{5'})^{1/2}\left(dr^2+r^2d\Omega_2^2\right)
\,,
\ena
where
\bea
h_{\tilde 2}&=&{t\over t_0}+{Q_{\tilde 2}\over r},
\nn
H_2&=&1+{Q_2\over r}~,~~
H_5=1+{Q_5\over r}~,~~
H_{5'}=1+{Q_{5'}\over r}
\,,
\ena
This metric is rewritten as
\bea
d\bar s_{4}^2&=&-(\tilde H_{\tilde 2}H_2H_5 H_{5'})^{-1/2} d\tau^2
+a_{\rm BH}^2(\tau)(\tilde H_{\tilde 2}H_2H_5 H_{5'})^{1/2}
\left(dr^2+r^2d\Omega_2^2\right)
\,,
\ena
where
\bea
\tilde H_{\tilde 2}=1+{\tilde{Q}_{\tilde 2}(\tau)\over r}
~,~~{\rm and}~~~
a_{\rm BH}=\left({\tau\over \tau_0}\right)^{1\over 3}
\,,
\label{p}
\ena
with
\bea
\tilde{Q}_{\tilde 2} \equiv \left({\tau\over \tau_0}\right)^{-{4\over 3}}
Q_{\tilde 2}
~,~~{\rm and}~~~
\tau_0 \equiv  {4\over 3}\, t_0
\,.
\ena
The power 1/3 in Eq.~(\ref{p}) is the one given in Table~\ref{table_BH}.

\section{Intersecting M-branes with M-waves and KK-monopoles}
\label{Mwave_KKmonopole}

Now we discuss the dynamical intersecting brane
solutions including M-waves and KK-monopoles in eleven dimensions.
The dimensional reduction of these generates the Kaluza-Klein
electric or magnetic charges in the 2-form field
strengths~\cite{Argurio:1998cp, Bergshoeff:1997tt, Sorkin:1983ns, Gross:1983hb}.
In $(D-1)$-dimensional spacetime, one can obtain the electric 0-brane
and the magnetic $(D-5)$-brane solutions.
Lifting up those solutions by one dimension, we obtain the KK-wave
and KK-monopole in $D$-dimensions, respectively.
In particular, KK-wave is called ``M-wave" in eleven-dimensional
theory~\cite{FigueroaO'Farrill:1999tx, FigueroaO'Farrill:2002tc}.
We briefly summarize those objects in Appendix \ref{KK_and_monopole}.

We extend our brane solutions given in \S~\ref{sec:classification}
to the cases with M-waves and/or KK-monopoles.
For the static case, there is a classification of the multiple intersecting
branes with the  M-waves and/or KK-monopoles~\cite{Bergshoeff:1996rn, Bergshoeff:1997tt}.

We first show the intersection rule for the branes with
M-wave and/or KK-monopoles, which is summarized in Table~\ref{rule_WKKM}.
In the Table, circles indicate where the brane world-volumes enter,
$\zeta$ represents the coordinate of the KK-monopole,
and the time-dependent branes are indicated by (a) and (b)
for different solutions.
When the solutions can be used for cosmology and black hole physics,
they are marked in the corresponding columns.

\begin{table}[ht]
\caption{The brane configurations
following the intersection rule with M-wave (W)
and/or KK-monopole (KKM).
The brane systems marked in the columns ``cos" and ``BH"
can be used for cosmological and black hole systems.
The labelling (a), (b), $\cdots$ in the column ``$\tilde I$"
denotes which brane (or M-wave, KK-monopole) is time dependent.
In the second case of M5-KKM system, there are two possibilities
which space dimensions can be our three space,
i.e., the case 1:
[$(\xi^1, \xi^2, \xi^3)
=(x^1, x^2, x^3)$] and the case 2:
[$(\xi^1, \xi^2, \xi^3)
=(x^4, x^5, x^6)$].
We show them by (c)-1
(c)-2, or (d)-1, (d)-2.
}
\label{rule_WKKM}
\begin{center}
{\scriptsize
\begin{tabular}{|c||c|c|c|c|c|c|c|c|c|c|c|c||c||c|c|}
\hline
&&0&1&2&3&4&5&6&7&8&9&10&$\tilde{I}$&cos&BH\\
\hline
M2-W&M2 & $\circ$ & $\circ$ & $\circ$ &&&&&&&&&(a)&$-$&$\surd$ \\
\cline{3-13}
&W & $\circ$ & $\zeta$ &&& & & & &&&&(b)&$-$&$\surd$\\
\hline
M5-W&M5 & $\circ$ & $\circ$ & $\circ$ & $\circ$ & $\circ$ & $\circ$ &&& & &
&(a)&$\surd$&$\surd$\\
\cline{3-13}
&W & $\circ$ & $\zeta$ &&& & & & &&&&(b)&$\surd$&$\surd$\\
\hline
M2-KKM&M2 & $\circ$ & $\circ$ & $\circ$ &&&&&&&&&(a)&$\surd$&$\surd$ \\
\cline{3-13}
&KKM & $\circ$ & $\circ$ & $\circ$ & $\circ$ & $\circ$ & $\circ$ & $\circ$ &
$\zeta$ & ${\cal A}^{\rm (m)}_8$ & ${\cal A}^{\rm (m)}_9$
& ${\cal A}^{\rm (m)}_{10}$&(b)&$\surd$&$\surd$\\
\cline{2-16}
&M2 & $\circ$ &&&&&&& $\circ$ &&& $\circ$&(c)&$\surd$&$-$  \\
\cline{3-13}
&KKM & $\circ$ & $\circ$ & $\circ$ & $\circ$ & $\circ$ & $\circ$ & $\circ$ &
$\zeta$ & ${\cal A}^{\rm (m)}_8$ & ${\cal A}^{\rm (m)}_9$
& ${\cal A}^{\rm (m)}_{10}$&(d)&$\surd$&$-$\\
\hline
M5-KKM&M5 & $\circ$ & $\circ$ & $\circ$ & $\circ$ & $\circ$ & $\circ$ & & &&&
&(a)&$\surd$&$\surd$\\
\cline{3-13}
&KKM& $\circ$ & $\circ$ & $\circ$ & $\circ$ & $\circ$ & $\circ$ & $\circ$ &
$\zeta$ & ${\cal A}^{\rm (m)}_8$ & ${\cal A}^{\rm (m)}_9$
& ${\cal A}^{\rm (m)}_{10}$&(b)&$\surd$&$\surd$\\
\cline{2-16}
&M5 & $\circ$ & $\circ$ & $\circ$ & $\circ$ &&& & $\circ$ &&& $\circ$
&(c)-1,2&$\surd$&$-$ \\
\cline{3-13}
&KKM& $\circ$ & $\circ$ & $\circ$ & $\circ$ & $\circ$ & $\circ$ & $\circ$ &
$\zeta$ & ${\cal A}^{\rm (m)}_8$ & ${\cal A}^{\rm (m)}_9$
& ${\cal A}^{\rm (m)}_{10}$&(d)-1,2&$\surd$&$-$\\
\hline
W-KKM&W
& $\circ$ & $\zeta^1$ &&& & & & &&&&(a)&$\surd$&$\surd$
\\
\cline{3-13}
&KKM
& $\circ$ & $\circ$ & $\circ$ & $\circ$ & $\circ$ & $\circ$ & $\circ$ &
$\zeta^7$ & ${\cal A}^{\rm (m)}_8$ & ${\cal A}^{\rm (m)}_9$
& ${\cal A}^{\rm (m)}_{10}$&(b)&$\surd$&$\surd$\\
\hline
KKM-KKM&KKM
& $\circ$ & $\circ$ & $\circ$ & $\circ$ & $\circ$ & ${\cal B}^{\rm (m)}_5$
 & ${\cal B}^{\rm (m)}_6$ &
$\zeta$ & $\circ$ & $\circ$ & ${\cal B}^{\rm (m)}_{10}$&(a)&$\surd$&$-$\\
\cline{3-13}
&KKM
& $\circ$ & $\circ$ & $\circ$ & $\circ$ & $\circ$ & $\circ$ & $\circ$ &
$\zeta$ & ${\cal A}^{\rm (m)}_8$ & ${\cal A}^{\rm (m)}_9$
& ${\cal A}^{\rm (m)}_{10}$&&&\\
\cline{2-16}
&KKM
& $\circ$ & $\circ$ & $\circ$ & $\circ$ & $\circ$ & $\zeta^5$
& ${\cal B}^{\rm (m)}_6$ &
$\circ$ & $\circ$ & ${\cal B}^{\rm (m)}_9$
& ${\cal B}^{\rm (m)}_{10}$&(b)&$\surd$&$-$\\
\cline{3-13}
&KKM
& $\circ$ & $\circ$ & $\circ$ & $\circ$ & $\circ$ & $\circ$ & $\circ$ &
$\zeta^7$ & ${\cal A}^{\rm (m)}_8$ & ${\cal A}^{\rm (m)}_9$
& ${\cal A}^{\rm (m)}_{10}$&& & \\
\hline
\end{tabular}
}
\end{center}
\end{table}
There are two configurations for two KK-monopole system as shown in Table
\ref{rule_WKKM}.
The metric of the former and the latter cases are given by
\Eqr{
&&ds_{\rm ~2KKM}^2
=-dt^2+\sum_{\alpha=1}^4(dx^{\alpha})^2
+h_{\rm m 1}\sum_{\alpha=5}^6(dz^\alpha)^2
+h_{\rm m 2}\sum_{\alpha=8}^9(dz^\alpha)^2
+h_{\rm m 1}h_{\rm m 2}\left(dz^{10}\right)^2\nn
&&~~
+(h_{\rm m 1}h_{\rm m 2})^{-1}\left[d\zeta+{\cal B}^{\rm (m)}_5dz^5
+{\cal B}^{\rm (m)}_6dz^6
+{\cal A}^{\rm (m)}_8dz^8+{\cal A}^{\rm (m)}_9dz^9
+\left({\cal A}^{\rm (m)}_{10}+{\cal B}^{\rm (m)}_{10}\right)dz^{10}
\right]^2.\nn
&&~~
  \label{rule:KMKM1 metric:Eq}
\\
&&ds_{\rm ~2KKM}^2
=-dt^2+\sum_{\alpha=1}^4(dx^{\alpha})^2+h_{\rm m 2}\left(dz^6\right)^2
+h_{\rm m 1}\left(dz^8\right)^2+h_{\rm m 1}h_{\rm m 2}
\sum_{\alpha=9}^{10}(dz^\alpha)^2\nn
&&~~
+(h_{\rm m2})^{-1}\left(d\zeta^5+{\cal B}^{\rm (m)}_6dz^6
+{\cal B}^{\rm (m)}_9dz^9+{\cal B}^{\rm (m)}_{10}dz^{10}\right)^2\nn
&&~~
+(h_{\rm m1})^{-1}\left(d\zeta^7+{\cal A}^{\rm (m)}_8dz^8
+{\cal A}^{\rm (m)}_9dz^9+{\cal A}^{\rm (m)}_{10}dz^{10}\right)^2\,.
 \label{rule:KMKM2 metric:Eq}
}

Next, we present the brane systems with M-wave or one KK-monopole.
As we mentioned, only one brane can have time dependence
in the present approach.
It is also true for the warp factor from M-wave or KK-monopole.
Hence we have two cases for time-dependent solutions, i.e.
we can have either one time-dependent brane or time-dependent M-wave (or KK-monopole).

In the former case, the metric forms for
the spacetimes with M-wave and with KK-monopole, respectively,
are written as
\bea
ds_{\rm W}^2&=& {\cal A}(t,z)
\Big[g_0(t,z)\left\{-dt^2+(d\zeta^{\tilde 1})^2+f_{\rm w}(z)
(dt-d\zeta^{\tilde 1})^2\right\}
\nn
&&
+\sum_{\tilde \alpha\neq 1, \tilde \alpha \in \tilde I}
g_{\tilde \alpha}(t,z)(dx^{\tilde \alpha})^2
+\sum_{\alpha \in \hspace{-.4em}/ \tilde I}
g_{\alpha}(z)(dy^{\alpha})^2+u_{ij}(z)dz^idz^j
\Big],
\\
ds_{\rm KKM}^2
&=& {\cal A}(t,z)
\Big[-g_0(t,z)dt^2+
\sum_{\tilde \alpha \in \tilde I}
g_{\tilde \alpha}(t,z)(dx^{\tilde \alpha})^2
\nn
&&
+\sum_{\alpha \in \hspace{-.4em}/ \tilde I}
g_{\alpha}(z)(dy^{\alpha})^2+
h_{\rm m}(z)u_{ij}dz^idz^j
+h_{\rm m}^{-1}(z)\left(d\zeta+{\cal A}_i^{\rm (m)}(z)dz^i\right)^2
\Big],
\label{metric_Mbrane+M}
\ena
with
\bea
{\cal A}&=&\left[A_{\tilde I}t+H_{\tilde I}(z)\right]^{a_{\tilde I}}
\prod_{I\neq \tilde I}H_I(z)^{a_{I}}~~,~~~
g_0\,=\, \left[A_{\tilde I}t+H_{\tilde I}(z)\right]^{-1}
\prod_{I\neq \tilde I}H_I(z)^{-1}, \nn
g_{\tilde \alpha}&=&
\left[A_{\tilde I}t+H_{\tilde I}(z)\right]^{-1}
\prod_{I\neq \tilde I}
H_I^{-\gamma_I^{(\tilde \alpha)}}~~,~~~
g_{\alpha} \,= \,
\prod_{I\neq \tilde I}H_I^{-\gamma_I^{(\alpha)}}
\,.
\ena
where
\bea
a_{\tilde I}={p_{\tilde I}+1\over D-2}~~\,,
~~{\rm and}~~~
\gamma^{(\alpha)}_I=\left\{
\begin{array}{cc}
1&~{\rm for}~~\alpha\in I~ \\ 0 &~{\rm for}~~\alpha
\in \hspace{-.8em}/ I \,.
\end{array} \right.
\ena
The coordinate $\zeta$ belongs to either $\Xsp$ or $\Zsp$.
In the KK-monopole case, dim($\Zsp$)=3.

If M-wave or KK-monopole depends on time,
we find the following solutions:
\bea
ds_{\rm  W }^2&=& {\cal A}(z)
\left[g_0(z)\left\{-dt^2+(d\zeta^{\tilde 1})^2+f_{\rm w}(t,z)
(dt-d\zeta^{\tilde 1})^2\right\}
\right.
\nn
&&
\left.
+\sum_{\alpha}
g_{\alpha}(z)(dx^{\alpha})^2
+u_{ij}(z)dz^idz^j
\right],
\\
~
ds_{\rm  KKM}^2
&=& {\cal A}(z)
\Bigl[-g_0(z)dt^2+
\sum_{\alpha }
g_{\alpha}(z)(dx^{\alpha})^2
\nn
&&
+
h_{\rm m}(t,z)u_{ij}dz^idz^j
+h_{\rm m}^{-1}(t,z)\left(d\zeta+{\cal A}^{\rm (m)}_i(z)dz^i
\right)^2 ~
\Bigr],
\ena
with
\bea
{\cal A}=\prod_{I}H_I^{a_I}~,~~
g_0=\prod_{I}H_I^{-1}~,~~
g_{\alpha}=
\prod_{I}H_I^{-\gamma_I^{(\alpha)}}
\,,
\ena
where
\bea
f_{\rm w}(t, z)=A_{\rm w} t+H_{\rm w}(z)-1~,~~
h_{\rm m}(t, z)=A_{\rm m} t+H_{\rm m}(z)
\,.
\ena
$H_{\rm w}$ and $H_{\rm m}$ are harmonic functions on $\Zsp$ space, and
${\cal A}_i^{\rm (m)}$ satisfies Eqs.~(\ref{KKM:potential:Eq}) and (\ref{KKM:A:Eq}).

We give one concrete example, i.e. the
M2-M5 brane system with M-wave (M2-M5-W).

\begin{table}[ht]
\label{M2M5W}
\caption{M2-M5-W brane system}
\begin{center}
{\scriptsize
\begin{tabular}{|c|c|c|c|c|c|c|c|c|c|c|c||c||c|c|}
\hline
&0&1&2&3&4&5&6&7&8&9&10&$\tilde I$&cos&BH
\\
\hline
M2 & $\circ$ & $\circ$ & $\circ$ &&&&&&&&&(a)&$\surd$& $\surd$
\\
\cline{2-15}
M5 & $\circ$ & $\circ$ &  & $\circ$ & $\circ$ & $\circ$
& $\circ$ & & & &&(b)&$\surd$&$\surd$
\\
\cline{2-15}
W & $\circ$ & $\circ$ &  & & & & & & & &&(c)&$\surd$&$\surd$
\\
\hline
& $t$ & $\zeta^{\tilde 1}$ & $y^{2}$& $x^{\tilde 3}$ & $x^{\tilde 4}$
& $x^{\tilde 5}$& $x^{\tilde 6}$ & $z^1$& $z^2$&$z^3$&$z^4$&&&
\\
\hline
\end{tabular}
}
\end{center}
\end{table}

If M5 brane is time-dependent
(M2-M5-W (b): See Table \ref{M2M5W} for the configuration),
the metric is then given by
\begin{eqnarray}
\label{M2M5W:metric:Eq}
ds^2&=& h_{\tilde 5}^{2/3}H_2^{1/3}\Bigl[(h_{\tilde 5}H_2)^{-1}
\left\{-dt^2+(d\zeta^{\tilde 1})^2+f_{\rm w}(dt-d\zeta^{\tilde 1})^2\right\}
\nn
&&
\hskip 2cm
+H_2^{-1} (dy^{2})^2
+h_{\tilde 5}^{-1} \sum_{\tilde \alpha=3}^6(dx^{\tilde \alpha})^2
     +u_{ij}dz^idz^j\Bigr]
\,,
\end{eqnarray}
that is
\begin{eqnarray}
{\cal A}=h_{\tilde 5}^{2/3}H_2^{1/3}\,,~~
 g_0=(h_{\tilde 5}H_5)^{-1}\,,~~
g_{\tilde 2}=g_{\tilde 3}=g_{\tilde 4}
=g_{\tilde 5}=h_{\tilde 5}^{-1}
\,,~~
g_{2}=H_2^{-1},
\end{eqnarray}
where
\begin{eqnarray}
h_{\tilde 5}=A_{\tilde 5}t+H_{\tilde 5}(z)
\,.
\end{eqnarray}
The form field is given by
\begin{eqnarray}
F_{(4)}&=-&\ast_{\Zsp} \left(dh_{\tilde 5}\right)\wedge dy^{2}
+dH_2^{-1}\wedge dt\wedge d\zeta^{\tilde 1}\wedge dy^{2}\,,
\end{eqnarray}
where $\ast_{\Zsp}$ is the Hodge dual operator
in the four-dimensional ${\Zsp}$ space.

Since the classification of static solutions is given
in~\cite{Bergshoeff:1996rn, Bergshoeff:1997tt} and ours is basically
the same, we discuss only interesting cases here.
As we discussed in \S~\ref{application},
we can apply the present solutions to analyze
cosmology and black holes.
In order to discuss those subjects,
we need either an isotropic and homogeneous three space
in the brane world-volume
or three-dimensional (or higher-dimensional) vacuum $\Zsp$ space.
However, as seen in Table \ref{rule_WKKM},
a wave breaks the isotropy and homogeneity in one wave-propagating dimension,
and KK-monopoles not only give inhomogeneities but also fill branes
in many dimensions after compactifying $\zeta$-direction.
These facts make the application to our interesting subjects
more difficult as the number of KK-monopoles increases.
As a result, we find several examples in the system of small number of
branes, but few examples for the system of large number of branes.
In Appendix \ref{Intersecting_branes_with_MKK},
we present the brane configurations for those possible models

\subsection{Cosmology}

For zero M-brane or one M-brane systems, which we show in Table
\ref{rule_WKKM}, we can discuss cosmology for many cases.
The possible cosmological models are marked by $\surd$ in the column ``cos"
(M5-W, M2-KKM, M5-KKM, W-KKM, KKM-KKM brane systems).
For two M-brane system, the possible models are
M2-M5-W, M2-M5-KKM, M2-M5-W-KKM, M2-M2-KKM, M5-M5-KKM, M5-M5-KKM-KKM.
For more than two brane system with M-waves or KK-monopoles, we
have no interesting case.
Note that we have only three cases (M2-W, M5-W, and M2-M5-W)
in which spacetime is regular on the branes (at $\bm{z}=\bm{z}_k$).
For other configurations, the curvature diverges. One may need
some mechanism to avoid singularity if our world is confined on the brane.

In Table \ref{table_power_WKK},
we summarize the power exponent of the scale factor $a$
of the expanding universe when the brane is time-dependent.
For the case that the wave or KK-monopole is time-dependent,
we always find the same power exponents, i.e,,
$\beta_{\rm c}=1/3$ and $-1/3$ for the time-dependent wave and the
time-dependent KK-monopole, respectively.
In some cases [M2-KKM(c), M2-M5-KKM(d)-1,2, M2-M2-KKM(d),
and the time-dependent wave (M5-W(b), W-KKM(a), M2-M5-W(c),
M2-M5-W-KKM(c))], we find that the power exponent
of the scale factor is 1/3,
which is that of the expanding universe with stiff matter fluid.

\begin{table}[ht]
\caption{
The power exponent $\beta_{\rm c}$ of the scale factor $a$ of
possible $4$-dimensional cosmological model is given, i.e. $a
\propto \tau^\beta_{\rm c}$, where $\tau$ is the cosmic time.
The labelling (a), (b), $\cdots$ corresponds to the configuration given in
Tables~4 and 16.
}
\begin{center}
{\scriptsize
\begin{tabular}{|c||c||c|c|}
\hline
&branes&${\rm dim}(\Zsp)$
&$\beta_{\rm c}$
\\
\hline
\hline
&M5-W(a)& 5 & $-1/2$
\\
\cline{2-4}
&M5-KKM(a)& 3&0
\\
\cline{2-4}
&M5-KKM(c)-1& 3& 1/5
\\
\cline{2-4}
&M5-KKM(c)-2&3&0
\\
\cline{2-4}
&M2-M5-W(b)&4& $-1/5$
\\
\cline{2-4}
case 1
&M2-M5-KKM(b)& 3& 0
\\
\cline{2-4}
(${\tilde I}=$M5)
&M2-M5-KKM(e)-1& 3&1/5
\\
\cline{2-4}
&M2-M5-KKM(e)-2& 3& 0
\\
\cline{2-4}
&M2-M5-W-KKM(b)& 3& 0
\\
\cline{2-4}
&M5-M5-KKM(a)& 3& 0
\\
\cline{2-4}
&M5-M5-KKM(b)& 3& 1/5
\\
\cline{2-4}
&M5-M5-KKM-KKM(a)& 4&1/13
\\
\hline
\hline
&M2-KKM(a)&3 &0
\\
\cline{2-4}
&M2-KKM(c)& 3& 1/3
\\
\cline{2-4}
&M2-M5-W(a)& 4& $-1/5$
\\
\cline{2-4}
case 2
&M2-M5-KKM(a)& 3& $0$
\\
\cline{2-4}
(${\tilde I}=$M2)
&M2-M5-KKM(d)-1& 3&1/3
\\
\cline{2-4}
&M2-M5-KKM(d)-2& 3&1/3
\\
\cline{2-4}
&M2-M5-W-KK(a)&3 & 0
\\
\cline{2-4}
&M2-M2-KKM(c)& 3&1/3
\\
\cline{2-4}
&M2-M2-KKM(d)& 3&0
\\
\hline
\end{tabular}
}
\label{table_power_WKK}
\end{center}
\end{table}

We give a simple example of M2-M5-KKM(a).
The metric is given by
\bea
ds^2 &=& h_{\tilde 2}^{1/3} H_5^{2/3} \Big[ (h_{\tilde 2} H_5)^{-1} \left\{
- dt^2 + (dy^1)^2 \right\} + h_{\tilde 2}^{-1}(dy^2)^2
+ H_5^{-1} \sum_{\tilde \a=3}^6 (dx^{\tilde\a})^2 \nn
&& + h_m^{-1}(d\zeta+{\cal A}_i dz^i)^2 +h_m u_{ij} dz^i dz^j \Big].
\ena
The compactified metric in the Einstein frame is
\bea
ds_4^2 = h_m^{-1/2} [ - h_{\tilde 2}^{-1} dt^2 + d\bm \xi^2].
\ena
This gives $\b_{\rm c}=0$ in Table~\ref{table_power_WKK}.

For the case with the wave, we can smear some dimensions ($<$dim($\Zsp$))
just as in \S~\ref{application_cosmology}, and the result
is exactly the same as the case without the wave.

\subsection{Time-dependent black holes}

We can also discuss some black hole spacetime by compactifying
the brane world-volume as in \S~\ref{application_BH}.
Although the spacetime is time dependent, near-brane geometry is the same
as that of the static brane solution.
If we find the horizon at $|\bm{z}|=0$ for the static
brane solution, we obtain a black hole geometry by compactification.
As for the possible spacetime for a black hole (or object),
we summarize our result in
Table \ref{table_BH_WKK}.
\begin{table}[ht]
\caption{
The power exponent $\beta_{\rm BH}$ of the scale factor
$a_{\rm BH}$ of the asymptotic FRW universe for the
possible $4$-dimensional black hole spacetime is given, i.e. $a_{\rm BH}
\propto \tau^{\beta_{\rm BH}}$, where $\tau$ is the cosmic time.
The marked one in the column ``BH" has a finite horizon area, i.e,,
it has a regular horizon.
}
\begin{center}
{\scriptsize
\begin{tabular}{|c||c||c|c||c|}
\hline
&branes&$d$
&$\beta_{\rm BH}$ &BH
\\
\hline
\hline
&M5-W& 6 &1/5 &
\\
\cline{2-5}
case 1
&M5-KKM& 4&1/3 &
\\
\cline{2-5}
(${\tilde I}=$M5)
&M2-M5-W&5& 1/4&$\surd$
\\
\cline{2-5}
&M2-M5-KKM& 4& 1/3&
\\
\cline{2-5}
&M2-M5-W-KKM& 4&1/3 &$\surd$
\\
\hline
\hline
&M2-W&9 & 1/8&
\\
\cline{2-5}
&M2-KKM&4 &1/3 &
\\
\cline{2-5}
case 2
&M2-M5-W& 5& 1/4&$\surd$
\\
\cline{2-5}
(${\tilde I}=$M2)
&M2-M5-KKM& 4& 1/3 &
\\
\cline{2-5}
&M2-M5-W-KKM&4 & 1/3&$\surd$
\\
\cline{2-5}
&M2-M2-KKM&4 & 1/3&
\\
\hline
\hline
&M5-W& 6&1/5 &
\\
\cline{2-5}
case 3
&W-KKM& 4& 1/3&
\\
\cline{2-5}
(${\tilde I}=$W)
&M2-M5-W&5& 1/4&$\surd$
\\
\cline{2-5}
&M2-M5-W-KKM& 4& 1/3 &$\surd$
\\
\hline
\hline
&M5-KKM& 4& 1/3&
\\
\cline{2-5}
&W-KKM& 4& 1/3&
\\
\cline{2-5}
case 4
&M2-M5-KKM&4 & 1/3 &
\\
\cline{2-5}
(${\tilde I}=$KKM)
&M2-M5-W-KKM& 4& 1/3&$\surd$
\\
\cline{2-5}
&M2-M2-KKM&4 & 1/3&
\\
\hline
\end{tabular}
}
\label{table_BH_WKK}
\end{center}
\end{table}

We show one concrete example of M2-M5-W brane system.
The metric is given by Eq. (\ref{M2M5W:metric:Eq}).
Compactifying the brane world-volume, the 5-dimensional metric
in the Einstein frame is given by
\bea
d\bar s_{5}^2&=&-[h_{\tilde 5}H_2 (1+f_{\rm w})]^{-2/3} dt^2
+[h_{\tilde 5}H_2 (1+f_{\rm w})]^{1/3}\left(dr^2+r^2d\Omega_3^2\right)
\,,
\ena
where
\bea
h_{\tilde 5}={t\over t_0}+{Q_{\tilde 5}\over r^2}
~~,~~~H_2=1+{Q_2\over r^2}~~,~~~
f_{\rm w}={Q_{\rm w}\over r^2}
\,,
\ena

This metric is rewritten as
\bea
d\bar s_{5}^2&=&-[\tilde H_{\tilde 5}H_2 (1+f_{\rm w})]^{-2/3} d\tau^2
+a_{\rm BH}^2(\tau)[\tilde H_{\tilde 5}H_2  (1+f_{\rm w})]^{1/3}
\left(dr^2+r^2d\Omega_3^2\right)
\,,~~
\ena
where
\bea
\tilde H_{\tilde 5}=1+{\tilde{Q}_{\tilde 5}(\tau)\over r^2}
~,~~{\rm and}~~~
a_{\rm BH}=\left({\tau\over \tau_0}\right)^{1\over 4}
\,,
\ena
with
\bea
\tilde{Q}_{\tilde 5} \equiv \left({\tau\over \tau_0}\right)^{-{3\over 2}}
Q_{\tilde 5}
~,~~{\rm and}~~~
\tau_0 \equiv  {3\over 2}\, t_0
\,.
\ena
The expansion rate of this scale factor is the same as
that of the stiff-matter dominant universe in 5 dimensions.
Hence this solution is regarded as a five-dimensional black hole
in the expanding universe.

\section{Lorentz invariance and the lower-dimensional effective theory}
\label{effective theory}

When we discuss four-dimensional cosmology, we assume that our three space is
isotropic and homogeneous.
However, in that case, the time direction can be different from three spatial directions.
In fact if some branes are not filled in this three space,
the time direction which is filled by all branes is not the same as
three spatial directions.
When we perform a Lorentz transformation, three spatial directions are not equivalent.
For example, suppose we have M2-M5-M5 brane system.
There are two possible cosmological models:
the case 1 (M2-M5-M5 (b)) and the case 2 (M2-M5-M5 (a))
(see Tables \ref{table_power} and  \ref{threeM}).
We assume that we are living on three space
$\bm{\xi}=(x^3,x^4,x^5)$.
Then the four dimensional metric in the Einstein frame is
\bea
d\bar{s}_{4}^2
=
 F_{\tilde I}(z)
\left[
-f_0(z)dt^2+f_\xi(z) d\bm{\xi}^2
\right]
\,,
\ena
where
\bea
&F_{\tilde 2}=(H_5H_{5'})^{-1/2}~~,~~~
f_0(z)=f_\xi(z) =(H_5H_{5'})^{-1}~~~~~~~& \mbox{for ~M2-M5-M5 (a)}
\,,
\nn
&F_{\tilde 5}=H_5^{-1/2}~~,~~~
f_0(z)=(H_2H_5)^{-1}~~,~~~
f_\xi(z) =H_5^{-1}~~~& \mbox{for ~M2-M5-M5 (b)}
\,.~~
\ena
For M2-M5-M5 (a), we have
\bea
d\bar{s}_{4}^2
=
(H_5H_{5'})^{-3/2}\left(-dt^2+d\bm{\xi}^2\right)
\propto \eta_{\mu\nu}d\xi^\mu d\xi^\nu
\,,
\ena
where $\xi^\mu=(t,\bm{\xi})$.
This spacetime is Lorentz invariant.
On the other hand, for M2-M5-M5 (b), we find
\bea
d\bar{s}_{4}^2
=
H_5^{-3/2}\left(-H_2^{-1}dt^2+d\bm{\xi}^2\right)
\,.
\ena
When we perform a Lorentz transformation in the
$t$-$\xi$ plane ($\xi=\xi^1$);
\bea
\left(
\begin{array}{c}
 t'\\
\xi'
\\
\end{array}
\right)=
\left(
\begin{array}{c}
\gamma\left(t-V \xi\right)\\
\gamma\left(\xi-V t \right)
\\
\end{array}
\right)
\,,
\ena
where $V$ is the velocity of new inertia frame and $\gamma=(1-V^2)^{-1/2}$ is
its Lorentz factor,
we have
\bea
d\bar{s}_{4}^2
=
H_5^{-3/2}\left[-(dt')^2+(d\bm{\xi'})^2+(H_2^{-1}-1)\gamma^2(dt'+Vd\xi')^2
\right]
\,.
\label{Lorentz_breaking}
\ena
The last term in Eq. (\ref{Lorentz_breaking}) gives the breaking term
of Lorentz invariance. Hence the order of magnitude of
breaking the Lorentz invariance is
\bea
{\cal O} \left(H_2^{-1}-1\right)
\sim {\cal O}\left(\sum_k {Q_{2,k}\over |\bm{z}-\bm{z}_k|}
\right).
\ena

In order to keep the Lorentz invariance, we need
the condition of
$f_0=f_\xi$, which means that all branes except for one time-dependent brane
contain our three space $\Xi$.
Hence 4D universes constructed from M2-M5 (a), M5-M5, M5-M5-M5, and M2-M5-M5 (a)
have the Lorentz invariance.

We can extend this four-dimensional Minkowski space into
a curved space with general covariance.
We take the following metric just as in Appendix \ref{Appendix:p}:
\bea
ds^2={\cal A}\left[g_0 q_{\mu\nu}d\xi^\mu d\xi^\nu
+\sum_{\tilde \alpha \in \hspace{-.4em}/
 \Xi,\, \tilde \alpha \in \tilde I}
g_{\tilde I}(dx^{\tilde \alpha})^2
+\sum_{\alpha \in \hspace{-.4em}/ \tilde I}g_\alpha
(dy^\alpha)^2
+u_{ij}dz^idz^j\right]
\,,
\ena
where we assume that the four-dimensional metric $q_{\mu\nu}$ depends only
on $\xi^\mu$ and that all branes (or all except for one time-dependent brane)
fill our four-dimensional spacetime.
Inserting this metric form, we find the solution
\bea
&&
R_{\mu\nu}\left(\hat \Xi\right)=0~~,~~~R_{ij}(\Zsp)=0,
\nn
&&
h_{\tilde I}=K_{\tilde I}(\xi)+H_{\tilde I}(z)~~,~~~h_{I}=H_{I}(z)
~~{\rm for}~~I\neq \tilde I,
\nn
&&
D_{\mu}D_{\nu}K_{\tilde I}=0,
\nn
&&
\lap_{\Zsp}H_{\tilde I}=0~~,~~~
\lap_{\Zsp}H_I=0~~{\rm for}~~
I \neq \tilde I
\,,
\ena
where $\hat \Xi$ is our four-dimensional spacetime.

Here let us point out the important fact on the nature of the
dynamical solutions described above.
We often discuss the four-dimensional effective theories, which are
derived from eleven-dimensional supergravity with branes.
In these occasions, we assume that there exists a Lorentz invariance
in the limit of a flat Minkowski space,
write down the scalar curvature, and then integrate
the eleven-dimensional action over compactified extra dimensions
to discuss the four-dimensional effective theories.

However the dynamical solutions of the above type with the
warp factors in eleven dimensions which depend on time and $z^i$
are usually not solutions of the effective four-dimensional theories.
This is because they are genuinely $D$-dimensional so that
one can never neglect the dependence on the
bulk space ($\Zsp$) in the basic equations.
Hence the solutions of the effective theories are quite often
inconsistent with the basic equations in eleven dimensions.

Let us show one explicit example of M5-M5 brane system.
The four-dimensional effective action is obtained by dimensional reduction
from our eleven-dimensional action. Our solution is
\bea
ds^2&=& (h_{\tilde 5}H_5)^{-1/3}\left[q_{\mu\nu}(\hat \Xi)
d\xi^{\mu}d\xi^{\nu}
     +H_5 r_{\tilde \alpha \tilde \beta}(\Xsp)
dx^{\tilde \alpha}dx^{\tilde \beta}\right.\nn
     & &\left.+h_{\tilde 5} s_{\alpha\beta}(\Ysp)dy^\alpha dy^\beta
     +h_{\tilde 5} H_5 u_{ab}(\Zsp)dz^adz^b\right],\nn
F_{(4)}&=-&\ast_{\Zsp} \left(dh_{\tilde 5}\right)\wedge
dx^{\tilde \alpha}\wedge
dx^{\tilde \beta}-\ast_{\Zsp}\left(dH_5\right)\wedge
 dy^\alpha\wedge dy^\beta\,,
\ena
where $\ast_{\Zsp}$ is the Hodge dual operator
in the ${\Zsp}$ space, and $q_{\mu\nu}(\hat \Xi)$,
$r_{\tilde \alpha \tilde \beta}(\Xsp)$, $s_{\alpha\beta}(\Ysp)$
and $u_{ij}(\Zsp)$
are the metrics on our four-dimensional spacetime $\hat \Xi$,
two-dimensional space $\Xsp$, two-dimensional space $\Ysp$ and
the three dimensional transverse space $\Zsp$.
Taking into account the ansatz $h_{\tilde 5}=K_{\tilde 5}(\xi)+H_{\tilde 5}(z)$
and $H_5(z)$, we find the eleven-dimensional scalar curvature $R$ is
\Eqr{
R&=&(h_{\tilde 5}H_5)^{1/3}R(\hat \Xi)+(h_{\tilde 5}H_5)^{-2/3}
\left[h_{\tilde 5}R(\Xsp) +H_5R(\Ysp)+R(\Zsp)\right] \nn
&&-\frac{5}{3}(h_{\tilde 5}H_5)^{1/3}h_{\tilde 5}^{-1}
\Box_{\hat \Xi}K_{\tilde 5}
-\frac{4}{3}(h_{\tilde 5}H_5)^{-2/3}\left(h_{\tilde 5}^{-1}\lap_{\Zsp}h_{\tilde 5}
+H_5^{-1}\lap_{\Zsp}H_5\right)\nn
&&+\frac{15}{18}(h_{\tilde 5}H_5)^{-2/3}u^{ij}
\left(\pd_i\ln h_{\tilde 5}
\pd_j \ln h_{\tilde 5}+\pd_i\ln H_5\pd_j \ln H_5\right),
\label{M5M5:Ricci scalar:Eq}
}
where $\Box_{\hat \Xi}$ and $\lap_{\Zsp}$
are the D'Alembertian and Laplace operator for $\hat \Xi$-spacetime
and $\Zsp$-space, respectively.
Assuming Ricci flatness for $\Xsp$, $\Ysp$ and $\Zsp$ spaces,
and harmonicity for $H_{\tilde 5}$ and $H_{5}$
($\lap_{\Zsp}H_{\tilde 5}=\lap_{\Zsp}H_{5}=0$),
we get
\Eqr{
S=\frac{1}{2\tilde{\kappa}^2} \int_{\hat \Xi} H(\xi) R(\hat \Xi)
\ast_{\hat \Xi}{\bf 1}_{\hat \Xi},
   \label{M5M5:effective action:Eq}
}
where $\ast_{\hat \Xi}$ denotes the Hodge dual operator on $\hat \Xi$,
we have dropped the surface terms,
$\tilde{\kappa}\equiv (V_{\Xsp}V_{\Ysp}V_{0})^{-1/2}\kappa$,
and $H(\xi)$ is defined by
\Eq{
K(\xi)=K_{\tilde 5}(\xi)+ \bar{c};\quad
\bar{c}:=V_{0}^{-1}\int_{\Zsp}H_{\tilde 5}H_5
\ast_{\Zsp}{\bf 1}_{\Zsp},
\label{M5M5:4-dim field:Eq}
}
where $\ast_{\Zsp}$ represents the Hodge dual operator on $\Zsp$, and
$V_{\Xsp}$, $V_{\Ysp}$, $V_0$ are given by
\begin{eqnarray}
V_{\Xsp}\,=\,\int_{\Xsp}\ast_{\Xsp}{\bf 1}_{\Xsp}, ~~
V_{\Ysp}\,=\,
\int_{\Ysp}\ast_{\Ysp}{\bf 1}_{\Ysp}, ~~
V_0\,=\,\int_{\Zsp}H_5\ast_{\Zsp}{\bf 1}_{\Zsp}.
\end{eqnarray}
The four-dimensional field equations are then given by
\bea
& &\hspace{-1.6cm}
R_{\mu\nu}\left(\hat \Xi\right)=K^{-1}D_{\mu} D_{\nu} K,
\nn
& &\hspace{-1.6cm}\Box_{\hat \Xi} K=0.
\ena
If the four-dimensional spacetime $\hat \Xi$ is Ricci flat, these
equations reproduce the correct ones for $K_{\tilde 5}(x)=K-\bar{c}$
obtained from the eleven-dimensional theory before.
However, the Ricci flatness of $\hat \Xi$ is not
required in the present effective theory (\ref{M5M5:effective action:Eq})
unlike in the full
eleven-dimensional theory (\ref{gs:D-dim action:Eq}).
Hence, the class of solutions
obtained in the four-dimensional effective theory are much
larger than the higher-dimensional original theory~\cite{Kodama:2005cz}.
This makes the higher-dimensional solutions even more restrictive
than those of the four-dimensional effective equations.
This is because the information of the internal space
which gives constraints on the lower dimensions was lost
after compactifying the internal space.

We note that the effective theory has a modular invariance
similar to the no-flux case $F_4=0$.
In fact, by the conformal transformation
$ds^2_{\hat \Xi}=K^{-1}ds^2_{\bar \Xi}$,
\Eqref{M5M5:effective action:Eq} is expressed in terms of the
variables in the Einstein frame as
\bea
S=\frac{1}{2\tilde{\kappa}^2} \int_{\bar \Xi }
 \left[R(\bar \Xi)\ast_{\bar \Xi}{\bf 1}_{\bar \Xi}
- \frac{3}{2}d\varphi \wedge \ast_{\bar \Xi}d\varphi
 \right],
\label{M5M5:effective action E-frame:Eq}
\ena
where $R(\bar \Xi)$ is the scalar curvature with respect to
the metric $ds^2_{\bar \Xi}$, $\ast_{\bar \Xi}$ denotes the Hodge dual
operator on $\bar \Xi$, and $\varphi\equiv \sqrt{\frac{3}{2}}\ln K$.
The corresponding four-dimensional Einstein equations
in the Einstein frame and the field equation for $\varphi$
are given by
\bea
& &R_{\mu\nu}(\bar \Xi)=\bar{D}_{\mu}\varphi\,
   \bar{D}_{\nu}\varphi,\nn
& &\Box_{\bar \Xi}\varphi=0
\,.
\ena
It is clear that this action and the equations of motion
are invariant under the transformation
$\varphi\rightarrow -\varphi+\lambda$,
where $\lambda$ is an arbitrary
constant.

\section{Concluding remarks}
  \label{sec:Discussions}

In this paper, we have derived general intersecting dynamical brane solutions,
given the complete classification of the intersecting M-branes,
and discussed the dynamics of the higher-dimensional supergravity models
with applications to cosmology and black hole physics.
The solutions we have found are the spacetime-dependent solutions.
These solutions were obtained by replacing a time-independent
warp factor $h_{\tilde I}=H_{\tilde I}(z)$ of
a supersymmetric solution by a time-dependent function
$h_{\tilde I}=A_{\tilde I} t+H_{\tilde I}(z)$
\cite{Kodama:2005fz, Binetruy:2007tu}.
Our solutions can contain only one function depending on both time $t$
 and transverse space coordinates $z^i$.

Supposing that our universe stays at a constant position in the bulk space Z
($\bm{z}_k$),
we have shown that several four-dimensional effective theories on the branes
give four-dimensional Minkowski space or FRW universe.
The power of the scale factor, however, is too
small to give a realistic expansion law.
This means that we have to consider additional matter on the brane
in order to get a realistic expanding universe.
On the other hand,
we can also discuss time-dependent black hole spacetimes which approach
asymptotically the FRW universe, if we regard the bulk space as our universe.
The near horizon geometries of these black holes in the expanding universe
are the same as the static solutions. However the asymptotic structures are
completely different, giving the FRW universe with scale factors same as
the universe filled by stiff matter.

In the viewpoint of higher-dimensional theory, the dynamics
of four-dimensional background are given by the solution of
higher-dimensional Einstein equations.
For instance, in the black $p$-brane system,
the solution tells us that the $(p+1)$-dimensional spacetime $\Xsp$
is Ricci flat. The $(p+1)$-dimensional spacetime is then similar
to the Kasner solution for the $(p+1)$-dimensional background~\cite{Binetruy:2007tu}.
On the other hand, if we start from the lower-dimensional effective
theory for warped compactification, the solutions may not be allowed in
the higher-dimensional theory. We have shown that it is the case
in M5-M5 brane system. The same is true for M5-M5-M5 and D2-D6 brane systems
in ten-dimensional type IIA theories.
This is because the function of $z$
in the metric is integrated out in the lower-dimensional effective action.
Then, the information of the extra dimensions in the function $h$
of the metric will be lost by the compactification.
This result implies that we have to be careful
when we use a four-dimensional effective theory to analyse
the moduli stabilization problem and the cosmological
problems in the framework of warped compactification of
supergravity or M-theory~\cite{Kodama:2005cz} (see
also~\cite{Koyama:2006ni, Shiu:2008ry, Arroja:2006zz, Giddings:2005ff,
Frey:2008xw, Arroja:2008ma} for recent progress in the effective
theory for warped compactifications).

We have also noted that if the Lorentz invariance is not kept
on the lower-dimensional world-sheet, the lower-dimensional
effective action  cannot be written in the covariant form
for the lower-dimensional metric. Some of the four- or five-dimensional
effective theories in this paper thus have broken Lorentz
invariance on the world-sheet.
Although the examples considered in the present paper do not
provide realistic cosmological models, this feature may be
utilised to investigate a cosmological analysis in a
realistic higher-dimensional cosmological model.

As we stated above,
our solutions can contain only one function depending on both time and
transverse space coordinates, and this seems to be a limitation on the applications
of the solutions. Recent study of similar systems depending
on the light-cone coordinate and space shows that it is possible to
obtain solutions with more nontrivial dependence on spacetime
coordinates~\cite{MOTW}. It is interesting to study if similar more
general solutions can be obtained by relaxing some of our ans\"atze.
We hope to report on this subject in the near future.

\section*{Acknowledgments}

K.U. would like to thank H. Kodama, M. Sasaki, T. Okamura, K. Nakao
for continuing encouragement, and E. Bergshoeff, H. Ishihara, P. Orland
for discussions.
Part of this work was carried out while two of the authors (N.O. and K.U.)
were attending SI2008. We thank the organizers for their hospitality.
This work was supported in part by the Grant-in-Aid for
Scientific Research Fund of the JSPS Nos. 19540308, 20540283 and 06042,
the Japan-U.K. Research Cooperative Program, and
Grant-in-Aid for Young Scientists
(B) of JSPS Research No.~20740147.


\appendix

\section{Dynamical solution of a single $p$-brane}
\label{Appendix:p}

In this appendix, we briefly summarize the results
for the case of  a single dynamical $p$-brane~\cite{Binetruy:2007tu}.
We consider  a single $p$-brane in our action (\ref{gs:D-dim action:Eq})~\cite{Lu:1995cs}.
In what follows, we use the same notation for the
variables and parameters of this
single brane dropping the suffix $I$.

To solve the field equations (\ref{gs:Einstein equations:Eq}),
(\ref{gs:scalar field equation:Eq}), and (\ref{gs:gauge field equation:Eq}),
 we assume the $D$-dimensional metric
in the form
\Eq{
ds^2=h^a(x, z)u_{ij}(\Zsp)dz^idz^j+
h^b(x, z)q_{\mu\nu}(\Xsp)dx^{\mu}dx^{\nu},
 \label{p:metric of p-brane solution:Eq}
}
where $q_{\mu\nu}$ is a $(p+1)$-dimensional metric which
depends only on the coordinates $x^{\mu}\equiv
(t, x^\alpha)$
with $\alpha$ being the spatial coordinates of the brane,
and $u_{ij}$ is the $(D-p-1)$-dimensional metric which
depends only on the coordinates $z^i$.
The parameters $a$ and $b$ are given in Eqs.~(\ref{gs:paremeter:Eq}).
Note that in the case of interacting branes, we divide the coordinate
for branes into two parts;
the time coordinate $t$ and the spatial coordinates of brane
world-volume $x^\alpha$, and assume that the metric depends on
only $t$ and $z^i$, but not on $x^\alpha$.
The metric form~(\ref{p:metric of p-brane solution:Eq}) is
a straightforward generalization of the case of a static $p$-brane
system with a dilaton coupling~\cite{Lu:1995cs}.

We also assume that the scalar field $\phi$ and
the gauge field strength $F_{(p+2)}$ are given by
Eq. (\ref{gs:ansatz for fields:Eq})

With the above ansatz, the Einstein equations are given by
\Eqrsubl{p:components of p-brane Einstein equations:Eq}{
&&R_{\mu\nu}(\Xsp)-h^{-1}D_{\mu}D_{\nu} h -\frac{b}{2}h^{-1}
q_{\mu\nu}\left(\triangle_{\Xsp}h + h^{-1}\triangle_{\Zsp} h\right)=0,
 \label{p:p-brane Einstein equation-mu:Eq}\\
&&R_{ij}(\Zsp)-\frac{a}{2} u_{ij}\left(\triangle_{\Xsp} h
 +h^{-1}\triangle_{\Zsp}h \right)=0,
 \label{p:p-brane Einstein equation-ij:Eq}\\
&&\pd_{\mu}\pd_i h=0,
 \label{p:p-brane Einstein equation-mi:Eq}
}
where $D_{\mu}$ is the covariant derivative with respective to
the metric $q_{\mu\nu}$,
$\triangle_{\Xsp}$ and $\triangle_{\Zsp}$ are
the Laplace operators on the space of
${\rm \Xsp}$ and the space ${\rm \Zsp}$, and
$R_{\mu\nu}(\Xsp)$ and $R_{ij}(\Zsp)$ are the Ricci tensors
of the metrics $q_{\mu\nu}$ and $u_{ij}$, respectively.
{}From Eq.~\eqref{p:p-brane Einstein equation-mi:Eq},
the warp factor $h$ must be in the form
\Eq{
h(x, z)= K(x)+H(z).
  \label{p:form of warp factor:Eq}
}
With this form of $h$, the other components of
the Einstein equations~\eqref{p:p-brane Einstein equation-mu:Eq}
and \eqref{p:p-brane Einstein equation-ij:Eq}
are rewritten as
\Eqrsubl{p:components of p-brane Einstein equations2:Eq}{
&&R_{\mu\nu}(\Xsp)-h^{-1}D_{\mu}D_{\nu}K -\frac{b}{2}h^{-1}
q_{\mu\nu} \left( \triangle_{\Xsp} K
+h^{-1}\triangle_{\Zsp}H\right)=0,
   \label{p:p-brane Einstein equation-mu2:Eq}\\
&&R_{ij}(\Zsp)-\frac{a}{2}u_{ij}\left(\triangle_{\Xsp}K
+h^{-1}\triangle_{\Zsp}H\right)=0.
   \label{p:p-brane Einstein equation-ij2:Eq}
   }

Next we consider the gauge field.
Under the assumption \eqref{gs:ansatz for fields:Eq},
we find
\Eq{
dF_{(p+2)}=h^{-1}(2\pd_i \ln h \pd_j \ln h + h^{-1}\pd_i\pd_j h)
dz^i\wedge dz^j\wedge\Omega(\Xsp) =0.
}
Thus, the Bianchi identity is automatically satisfied.
Also the equation of motion for the gauge field
(\ref{gs:gauge field equation:Eq}) becomes
\Eqr{
d\left[e^{-c\phi}\ast F_{(p+2)} \right]
=-d\left[\pd_ih (\ast_{\Zsp}dz^i)\right]=0,
 }
where $\ast_{\Zsp}$ denotes the Hodge dual operator on $\Zsp$.
Hence, the gauge field equation is automatically
satisfied.

Let us consider the scalar field equation.
Substituting the forms of the scalar field
and the gauge field (Eq.~\eqref{gs:ansatz for fields:Eq}),
and the warp factor \eqref{p:form of warp factor:Eq}
into the equation of motion for the scalar field
\eqref{gs:scalar field equation:Eq}, we obtain
\Eq{
\frac{c}{2}h^{-a}\left(\triangle_{\Xsp}K
+h^{-1}\triangle_{\Zsp}H\right)=0,
  \label{p:p-brane scalar field equation2:Eq}
}
Thus, unless the parameter $c$ is zero, the warp factor $h$ should
satisfy the equations
\Eq{
\triangle_{\Xsp}K=0, ~~~ \triangle_{\Zsp}H=0.
   \label{p:solution for the scalar field:Eq}
}
If $F_{(p+2)}\ne 0$, the function $H$ is non-trivial.
In this case, the Einstein equations reduce to
\bea
&&R_{\mu\nu}(\Xsp)=0,\nn
&&R_{ij}(\Zsp)=0,\\
&&D_{\mu}D_{\nu}K=0.\nonumber
\ena
If $F_{(p+2)}=0$, however, the function $H$ becomes
trivial, and then  the internal space is no longer
warped~\cite{Kodama:2005cz}.

We show  an example. We consider the case
\Eq{
q_{\mu\nu}=\eta_{\mu\nu}\,,
\quad u_{ij}=\delta_{ij}\,,
 \label{p:special metric of p-brane solution:Eq}
 }
that is, we have  the $(p+1)$-dimensional
Minkowski space and the $(D-p-1)$-dimensional  Euclidean space.
In this case, the solution for $h$ is obtained explicitly as
\Eq{
h(x, z)=A_{\mu}x^{\mu}+B
+\sum_{k}\frac{Q_k}{|\bm{z}-\bm{z}_k|^{D-p-3}},
 \label{p:special form of warp factor for p-brane solution:Eq}
}
where $A_{\mu}$, $B$ and $Q_k$ are constant parameters.

For the case of $c=0$,
the scalar field becomes constant because of
the ansatz~\eqref{gs:ansatz for fields:Eq}, and
the scalar field equation \eqref{p:p-brane scalar field equation2:Eq}
is automatically satisfied.
Then, the Einstein equations become
\bea
&&R_{\mu\nu}(\Xsp)=0, \nn
&&R_{ij}(\Zsp)=\frac{1}{2}a (p+1)\,\lambda \, u_{ij}(\Zsp),\\
&&D_{\mu}D_{\nu}K=\lambda \, q_{\mu\nu}(\Xsp), \nonumber
\ena
where $\lambda$ is a constant.
We see that  the internal space $\Zsp$ is not Ricci flat,
but the Einstein space if $\lambda\neq 0$, and
the function $K$ can be more non-trivial.
For example, if  $q_{\mu\nu}=\eta_{\mu\nu}$, $K$
is no longer linear
 but quadratic  in the coordinates $x^\mu$~\cite{Kodama:2005fz}.

\section{Classification of Intersecting branes}
\label{Appendix:table_intersectingMbranes}

In this appendix, we present a complete classification of
time-dependent intersecting M-branes.
For two $\sim$ four brane systems,
we give all possible brane configurations and those metrics
explicitly in Tables \ref{table_2} -- \ref{table_4-2}.
In the first tables in these Tables, circles indicate where the brane
world-volumes enter, and the time-dependent branes are indicated by
(a) and (b) for different solutions.
When the solutions can be used for cosmology and black hole physics,
they are marked in the corresponding columns.
In the second (continued) tables, concrete metrics are given in the notation of \S~3
with the dimensions of transverse space Z for each time-dependent case
indicated in the first tables by (a) and (b).

For more than four branes, we show only simplified tables
(Tables~\ref{table_5} -- \ref{table_8}) to save the space because these
systems do not have applications to cosmology and black hole physics,
and are not so interesting.
They are included for the sake of completeness.
In these tables, we show which branes are involved, and dimension of the transverse
space Z, and the following columns with $k$M give the numbers of dimensions containing
$k$ branes. For example, (2, 2, 2, 2, 1) in the first row of Table~\ref{table_5}
means that there are these numbers of dimensions in which the world-volumes of
1 M-brane, 2 M-branes and so on lie.
Though these are not so explicit, they are useful to identify the explicit
brane configurations with higher numbers of branes from the systems with
lower numbers step by step.
In the next column is given how many different time-dependent solutions are
obtained according to which brane we give the time dependence.
For example, M5(3) in the first column of Table~\ref{table_5} means that
there are only three kinds of different solutions when we choose different
time-dependent M5 branes. This is because there are same kind of M5 branes
which give the same time-dependent solutions.
Which brane gives different time-dependent solutions can be easily
identified if we check the patterns of how many branes each coordinate
of the brane contains.

\begin{table}[h]
\caption{Intersections of two M-branes.}
\begin{center}
{\scriptsize
\begin{tabular}{|c||c|c|c|c|c|c|c|c|c|c|c|c||c||c|c|}
\hline
&&0&1&2&3&4&5&6&7&8&9&10&$\tilde{I}$&cos&BH
\\
\hline
(M2)$^2$&M2 & $\circ$ & $\circ$ & $\circ$ &  &   &   &&&&&& (a)
&$-$&$\surd$
\\
\cline{3-13}
&M2 & $\circ$ &   &   & $\circ$ &$\circ$&&  & & & &&&&
\\
\hline
\hline
M2M5&M2 & $\circ$ & $\circ$ &  &   & & & $\circ$ & & & &&(a)&$\surd$&$\surd$
\\
\cline{3-13}
&M5 & $\circ$ & $\circ$ & $\circ$ & $\circ$ &$\circ$ &$\circ$  &&&&&& (b)
&$\surd$&$\surd$
\\
\hline
\hline
(M5)$^2$
&M5 & $\circ$ & $\circ$ & $\circ$ & $\circ$ & $\circ$ & $\circ$ &&&&&& (a)
&$\surd$&$\surd$
\\
\cline{3-13}
&M5 & $\circ$ & $\circ$ & $\circ$ & $\circ$ &&& $\circ$ & $\circ$ & & &&&&
\\
\hline
\end{tabular}
}
\label{table_2}
\end{center}
\end{table}
\setcounter{table}{7}
\vs{-5}
\begin{table}[h]
\caption{(Continue) Concrete metrics for two M-branes.}
\begin{center}
{\scriptsize
\begin{tabular}{|c|c||c|c|}
\hline
\multicolumn{2}{|c||}{M2M2}&
${\cal A}=h_{\tilde 2}^{1/3}H_2^{1/3}$&
$g_{\tilde 0}=h_{\tilde 2}^{-1}H_2^{-1}$
\\
\hline
&${\rm dim}(\Zsp)$&
$g_{\tilde \alpha}$&$g_{\alpha}$
\\
\hline
(a)& 5
&$g_{\tilde 1}=g_{\tilde 2}=h_{\tilde 2}^{-1}$
&$g_{3}=g_{4}=H_2^{-1}$
\\
\hline
\hline
\multicolumn{2}{|c||}{M2M5}&
${\cal A}=h_{\tilde 2}^{1/3}H_5^{2/3}$&
$g_{0}=h_{\tilde 2}^{-1}H_5^{-1}$
\\
\hline
&${\rm dim}(\Zsp)$&
$g_{\tilde \alpha}$&$g_{\alpha}$
\\
\hline
(a)& 4
&$g_{\tilde 1}=h_{\tilde 2}^{-1}H_5^{-1},
g_{\tilde 6}=h_{\tilde 2}^{-1}$
&$g_{2}=g_{3}=g_{4}=g_{5}=H_5^{-1}$
\\
\hline
\hline
\multicolumn{2}{|c||}{M2M5}&
${\cal A}=h_{\tilde 5}^{2/3}H_5^{1/3}$&
$g_{0}=h_{\tilde 5}^{-1}H_5^{-1}$
\\
\hline
&${\rm dim}(\Zsp)$&
$g_{\tilde \alpha}$&$g_{\alpha}$
\\
\hline
(b)& 4
&$g_{\tilde 1}=h_{\tilde 5}^{-1}H_2^{-1}$
&$g_{6}=H_2^{-1}$
\\
 &
&$g_{\tilde 2}=g_{\tilde 3}=
g_{\tilde 4}=g_{\tilde 5}=h_{\tilde 5}^{-1}$
&
\\
\hline
\hline
\multicolumn{2}{|c||}{M5M5}&
${\cal A}=h_{\tilde 2}^{2/3}H_2^{2/3}$&
$g_{\tilde 0}=h_{\tilde 2}^{-1}H_2^{-1}$
\\
\hline
&${\rm dim}(\Zsp)$&
$g_{\tilde \alpha}$&$g_{\alpha}$
\\
\hline
(a)& 3
&$g_{\tilde 1}=g_{\tilde 2}=g_{\tilde 3}=
h_{\tilde 5}^{-1}H_2^{-1}$
&$g_{6}=g_{7}=H_5^{-1}$
\\
 &
&$g_{\tilde 4}=g_{\tilde 5}=h_{\tilde 2}^{-1}$
&
\\
\hline
\hline
\end{tabular}
}
\label{table_22}
\end{center}
\end{table}

\begin{table}[ht]
\caption{Intersections of three M-branes.}
\label{threeM}
{\scriptsize
\begin{center}
\begin{tabular}{|c||c|c|c|c|c|c|c|c|c|c|c|c||c||c|c|}
\hline
&&0&1&2&3&4&5&6&7&8&9&10&$\tilde{I}$&cos&BH
\\
\hline
&M5 & $\circ$ & $\circ$ & $\circ$ & $\circ$ & $\circ$ & $\circ$ &&&&&&(a)
&$\surd$&$-$
\\
\cline{3-13}
&M5 & $\circ$ & $\circ$ & $\circ$ & $\circ$ &&& $\circ$ & $\circ$ & & &&
&&
\\
\cline{3-13}
&M5 & $\circ$ &$\circ$&  $\circ$ & $\circ$ &&&  & &  $\circ$&$\circ$&&
&&
\\
\cline{2-16}
&M5 & $\circ$ & $\circ$ & $\circ$ & $\circ$ & $\circ$ &$\circ$ &&&&&&(b)
&$-$&$\surd$
\\
\cline{3-13}
(M5)$^3$
&M5 & $\circ$ & $\circ$ & $\circ$  &$\circ$&  && $\circ$ &  $\circ$&   & &&
&&
\\
\cline{3-13}
&M5 & $\circ$ &$\circ$&   &  &$\circ$&$\circ$& $\circ$ &$\circ$ &  &&&
&&
\\
\cline{2-16}
&M5 & $\circ$ & $\circ$ & $\circ$ & $\circ$ & $\circ$ &$\circ$ &&&&&&   (c)
&$-$&$-$
\\
\cline{3-13}
&M5 & $\circ$ & $\circ$ & $\circ$ & $\circ$ &&
& $\circ$ & $\circ$ & & &&
&&
\\
\cline{3-13}
&M5 & $\circ$ &$\circ$& $\circ$ &  &$\circ$&&  $\circ$& & $\circ$ &&&
&&
\\
\hline
\hline
&M2 & $\circ$ & $\circ$ & $\circ$ &  &   &  &&&&&& (a)
&$\surd$&$\surd$
\\
\cline{3-13}
&M5 & $\circ$ & $\circ$ &    &$\circ$& $\circ$ &$\circ$&  $\circ$ & &  & &&(b)
&$\surd$&$\surd$
\\
\cline{3-13}
M2(M5)$^2$
&M5 & $\circ$ & &  $\circ$ & $\circ$ &$\circ$&$\circ$&  & $\circ$ &  &&&
&&
\\
\cline{2-16}
&M2 & $\circ$ & $\circ$ & $\circ$ &   &   &   &&&&&&  (a)
&$-$&$-$
\\
\cline{3-13}
&M5 & $\circ$ &$\circ$ & & $\circ$ &$\circ$ & $\circ$& $\circ$ & & & && (b)
&$-$&$-$
\\
\cline{3-13}
&M5 & $\circ$ &$\circ$&   &  &&$\circ$& $\circ$ & $\circ$ & $\circ$ &&&
&&
\\
\hline
\hline
&M2 & $\circ$ & $\circ$ & $\circ$ &  &  & &&&&&& (a)
&$\surd$&$\surd$
\\
\cline{3-13}
(M2)$^2$M5
&M2 & $\circ$ &   & &$\circ$ & $\circ$ &  &  & & & &&
&&
\\
\cline{3-13}
&M5 & $\circ$ & $\circ$ & &$\circ$ &   & $\circ$& $\circ$ & $\circ$&  & &&
(b)&$\surd$&$\surd$
\\
\hline
\hline
&M2 & $\circ$ & $\circ$ & $\circ$ &   &   & & &&&&& (a)
&$-$&$\surd$
\\
\cline{3-13}
(M2)$^3$&M2 & $\circ$ &     &  & $\circ$ &$\circ$&
&   & & & &&  &&
\\
\cline{3-13}
&M2 &  $\circ$ & &     &   &  &$\circ$& $\circ$ & &   &&&
&&
\\
\hline
\end{tabular}
\end{center}
}
\label{table_3}
\end{table}
\setcounter{table}{8}
\begin{table}[ht]
\caption{(Continue) Concrete metrics for three M-branes.}
\begin{center}
{\scriptsize
\begin{tabular}{|c|c||c|c|}
\hline
\multicolumn{2}{|c||}{M5$^3$}&
${\cal A}=h_{\tilde 5}^{2/3}H_5^{2/3}H_{5'}^{2/3}$&
$g_{\tilde 0}=h_{\tilde 5}^{-1}H_5^{-1}H_{5'}^{-1}$
\\
\hline
&${\rm dim}(Z)$&
$g_{\tilde \alpha}$&$g_{\alpha}$
\\
\hline
(a)& 1
&$g_{\tilde 1}=g_{\tilde 2}=g_{\tilde 3}=h_{\tilde 5}^{-1}H_5^{-1}H_{5'}^{-1}$
&$g_{6}=g_{7}=H_5^{-1}$
\\
  &
&$g_{\tilde 4}=g_{\tilde 5}=h_{\tilde 5}^{-1}$
&$g_{8}=g_{9}=H_{5'}^{-1}$
\\
\hline
(b)& 3
&$g_{\tilde 1}=h_{\tilde 5}^{-1}H_5^{-1}H_{5'}^{-1}$
&
\\
  &
&$g_{\tilde 2}=g_{\tilde 3}=h_{\tilde 5}^{-1}H_5^{-1}$
&$g_{6}=g_{7}=H_5^{-1}H_{5'}^{-1}$
\\
  &
& $g_{\tilde 4}=g_{\tilde 5}=h_{\tilde 5}^{-1}H_{5'}^{-1}$
&
\\
\hline
(c)& 2
&$g_{\tilde 1}=g_{\tilde 2}=h_{\tilde 5}^{-1}H_5^{-1}H_{5'}^{-1}$
&$ g_{6}=H_5^{-1}H_{5'}^{-1}$
\\
  &
&$g_{\tilde 3}=h_{\tilde 5}^{-1}H_5^{-1},
g_{\tilde 4}=h_{\tilde 5}^{-1}H_{5'}^{-1}$
&$g_{7}=H_5^{-1}$
\\
  &
& $ g_{\tilde 5}=h_{\tilde 5}^{-1}$
&$g_{8}=H_{5'}^{-1}$
\\
\hline
\hline
\multicolumn{2}{|c||}{M2M5$^2$}&
${\cal A}=h_{\tilde 2}^{1/3}H_5^{2/3}H_{5'}^{2/3}$&
$g_{0}=h_{\tilde 2}^{-1}H_5^{-1}H_{5'}^{-1}$
\\
\hline
&${\rm dim}(Z)$&
$g_{\tilde \alpha}$&$g_{\alpha}$
\\
\hline
(a)& 3
&$g_{\tilde 1}=h_{\tilde 2}^{-1}H_5^{-1}$
&$g_{3}=g_{4}=g_{5}=H_5^{-1}H_{5'}^{-1}$
\\
 &
&$g_{\tilde 2}=h_{\tilde 2}^{-1}H_{5'}^{-1}$
&$g_{6}=H_5^{-1}, g_{7}=H_{5'}^{-1}$
\\
\hline
(b)& 2
&$g_{\tilde 1}=h_{\tilde 2}^{-1}H_5^{-1}H_{5'}^{-1}$
&$g_{3}=g_{4}=H_5^{-1}$
\\
 &
&$g_{\tilde 2}=h_{\tilde 2}^{-1}$
&$g_{5}=g_{6}=H_5^{-1}H_{5'}^{-1}$
\\
 &
&
&$g_{7}=g_{8}=H_{5'}^{-1}$
\\
\hline
\hline
\multicolumn{2}{|c||}{M2M5$^2$}&
${\cal A}=h_{\tilde 5}^{2/3}H_2^{1/3}H_{5}^{2/3}$&
$g_{0}=h_{\tilde 5}^{-1}H_2^{-1}H_{5}^{-1}$
\\
\hline
&${\rm dim}(Z)$&
$g_{\tilde \alpha}$&$g_{\alpha}$
 \\
\hline
(c)& 3
&$g_{\tilde 1}=h_{\tilde 5}^{-1}H_2^{-1}$
&$g_{2}=H_2^{-1}H_5^{-1}$
\\
&
&$g_{\tilde 3}=g_{\tilde 4}=g_{\tilde 5}=h_{\tilde 5}^{-1}H_{5}^{-1}$
&$g_{7}=H_{5}^{-1}$
\\
&
&$g_{\tilde 6}=h_{\tilde 5}^{-1}$
&
\\
\hline
(d)& 2
&$g_{\tilde 1}=h_{\tilde 5}^{-1}H_2^{-1}H_{5}^{-1}$
&$g_{2}=H_2^{-1}$
\\
 &
&$g_{\tilde 3}=g_{\tilde 4}=h_{\tilde 5}^{-1}$
&$g_{7}=g_{8}=H_{5}^{-1}$
\\
 &
&$g_{\tilde 5}=g_{\tilde 6}=h_{\tilde 5}^{-1}H_5^{-1}$
&
\\
\hline
\hline
\multicolumn{2}{|c||}{M2$^2$M5}&
${\cal A}=h_{\tilde 2}^{1/3}H_2^{1/3}H_5^{2/3}$&
$g_{0}=h_{\tilde 2}^{-1}H_2^{-1}H_5^{-1}$
\\
\hline
&${\rm dim}(Z)$&
$g_{\tilde \alpha}$&$g_{\alpha}$
\\
\hline
(a)& 3
&$g_{\tilde 1}=h_{\tilde 2}^{-1}H_5^{-1}$
&$g_{3}=H_2^{-1}H_5^{-1}$
\\
 &
&$g_{\tilde 2}=h_{\tilde 2}^{-1}$
&$g_{4}=H_2^{-1}$
\\
 &
&$g_{\tilde 1}=h_{\tilde 5}^{-1}H_2^{-1}$
&
\\
\hline
\multicolumn{2}{|c||}{M2$^2$M5}&
${\cal A}=h_{\tilde 5}^{2/3}H_2^{1/3}H_{2'}^{1/3}$&
$g_{0}=h_{\tilde 5}^{-1}H_2^{-1}H_{2'}^{-1}$
\\
\hline
&${\rm dim}(Z)$&
$g_{\tilde \alpha}$&$g_{\alpha}$
\\
\hline
(b)& 3
&$g_{\tilde 1}=h_{\tilde 5}^{-1}H_2^{-1}$
&
\\
 &
&$g_{\tilde 3}=h_{\tilde 5}^{-1}H_{2'}^{-1}$
&$g_{2}=H_2^{-1}, g_{4}=H_{2'}^{-1}$
\\
 &
&$g_{\tilde 5}=g_{\tilde 6}=g_{\tilde 7}=h_{\tilde 5}^{-1}$
&
\\
\hline
\hline
\multicolumn{2}{|c||}{M2$^3$}&
${\cal A}=h_{\tilde 2}^{1/3}H_2^{1/3}H_{2'}^{1/3}$&
$g_{0}=h_{\tilde 2}^{-1}H_2^{-1}H_{2'}^{-1}$
\\
\hline
&${\rm dim}(Z)$&
$g_{\tilde \alpha}$&$g_{\alpha}$
\\
\hline
(a)& 4
&$g_{\tilde 1}=g_{\tilde 2}=h_{\tilde 2}^{-1}$
&$g_{3}=g_{4}=H_{2}^{-1}$
\\
 &
&
&$g_{5}=g_{6}=H_{2'}^{-1}$
\\
\hline
\end{tabular}
}
\label{table_32}
\end{center}
\end{table}

\begin{table}[ht]
\caption{Intersections of four M-branes I.}
{\scriptsize
\begin{center}
\begin{tabular}{|c||c|c|c|c|c|c|c|c|c|c|c|c||c||c|c|}
\hline
&&0&1&2&3&4&5&6&7&8&9&10&$\tilde{I}$&cos&BH
\\
\hline
&M5 & $\circ$ & $\circ$ & $\circ$ & $\circ$ & $\circ$ & $\circ$ &&&&&& (a)
&$-$&$-$
\\
\cline{3-13}
&M5 & $\circ$ & $\circ$ & $\circ$ & $\circ$ &&& $\circ$ & $\circ$ & & &&(b)
&$-$&$-$
\\
\cline{3-13}
&M5 & $\circ$ & $\circ$ & $\circ$  && $\circ$ &&  $\circ$ & & $\circ$& &&
&&
\\
\cline{3-13}
&M5 & $\circ$ &$\circ$&   & $\circ$ &$\circ$ &&   &$\circ$ & $\circ$&&&
& &
\\
\cline{2-16}
&M5 & $\circ$ & $\circ$ & $\circ$ & $\circ$ & $\circ$ & &$\circ$&&&&& (c)
&$-$&$-$
\\
\cline{3-13}
&M5 & $\circ$ & $\circ$ & $\circ$ & $\circ$ && $\circ$
&  & $\circ$ & & &&&&
\\
\cline{3-13}
&M5 & $\circ$ & $\circ$ & $\circ$  && $\circ$ & $\circ$&  & & $\circ$ & &&&&
\\
\cline{3-13}
(M5)$^4$
&M5 & $\circ$ &$\circ$&   & $\circ$ &$\circ$&$\circ$&  & &  &$\circ$&&&&
\\
\cline{2-16}
&M5 & $\circ$ & $\circ$ & $\circ$ & $\circ$ & $\circ$ & $\circ$&&&&&&(d)
&$-$&$-$
\\
\cline{3-13}
&M5 & $\circ$ & $\circ$ & $\circ$ & $\circ$ &&
& $\circ$ & $\circ$ & & &&(e)
&$-$&$-$
\\
\cline{3-13}
&M5 & $\circ$ & $\circ$ & $\circ$  && $\circ$ & & $\circ$ & & $\circ$ & &&&&
\\
\cline{3-13}
&M5 & $\circ$ &$\circ$&  $\circ$ &  &&$\circ$& $\circ$ & &  &$\circ$&&&&
\\
\cline{2-16}
&M5 & $\circ$ & $\circ$ & $\circ$ & $\circ$ & $\circ$ & $\circ$&&&&&& (f)
&$-$&$-$
\\
\cline{3-13}
&M5 & $\circ$ & $\circ$ & $\circ$ & $\circ$ &&
& $\circ$ & $\circ$ & & &&&&
\\
\cline{3-13}
&M5 & $\circ$ & $\circ$ & $\circ$  && $\circ$ &&  $\circ$ & & $\circ$ & &&&&
\\
\cline{3-13}
&M5 & $\circ$ &$\circ$&  $\circ$ &  &&$\circ$&  &$\circ$ & $\circ$ &&&&&
\\
\hline
\hline
&M2 & $\circ$ & $\circ$ & $\circ$ &   &   &   &&&&&&  (a)
&$-$&$-$
\\
\cline{3-13}
&M5 & $\circ$ & $\circ$ & & $\circ$ &$\circ$ &$\circ$ & $\circ$ & & & && (b)
&$-$&$-$
\\
\cline{3-13}
&M5 & $\circ$ & $\circ$ &   &$\circ$& $\circ$ & & & $\circ$ &$\circ$ & &&&&
\\
\cline{3-13}
&M5 & $\circ$ &$\circ$&   &  &&$\circ$& $\circ$ & $\circ$&  $\circ$&&& &&
\\
\cline{2-16}
&M2 & $\circ$ & $\circ$ & $\circ$ &  &  & &&&&&& (c)
&$-$&$-$
\\
\cline{3-13}
&M5 & $\circ$ & $\circ$ &  & $\circ$ &$\circ$&
& $\circ$ &  $\circ$ & & &&(d)
&$-$&$-$
\\
\cline{3-13}
M2
(M5)$^3$
&M5 & $\circ$ & $\circ$ & &$\circ$ & $\circ$ & $\circ$&  & & $\circ$ & &&&&
\\
\cline{3-13}
&M5 & $\circ$ &$\circ$&   & $\circ$ &&$\circ$& $\circ$ & &  &$\circ$&&&&
\\
\cline{2-16}
&M2 & $\circ$ & $\circ$ & $\circ$ &   &   & & &&&&&(e)
&$-$&$-$
\\
\cline{3-13}
&M5 & $\circ$ & $\circ$   &  & $\circ$ &$\circ$&$\circ$ &
& $\circ$ & & & &(f)
&$-$&$-$
\\
\cline{3-13}
&M5 &  $\circ$ &  $\circ$ &    &$\circ$ & $\circ$ && $\circ$  &  &   $\circ$
  & &&&&
\\
\cline{3-13}
&M5 &  $\circ$ & &  $\circ$  & $\circ$ &$\circ$ &$\circ$& $\circ$  & &  &&&(g)
&$-$&$-$
\\
\hline
\end{tabular}
\end{center}
}
\label{table_4-1}
\end{table}
\setcounter{table}{9}
\begin{table}[ht]
\caption{(Continue) Concrete metrics for four M-branes I.}
\begin{center}
{\scriptsize
\begin{tabular}{|c|c||c|c|}
\hline
\multicolumn{2}{|c||}{M5$^4$}&
${\cal A}=(h_{\tilde 5}H_5H_{5'}H_{5''})^{2/3}$&
$g_{\tilde 0}=(h_{\tilde 5}H_5H_{5'}H_{5''})^{-1}$
\\
\hline
&${\rm dim}(Z)$&
$g_{\tilde \alpha}$&$g_{\alpha}$
\\
\hline
(a)& 2
&$g_{\tilde 1}=(h_{\tilde 5}H_5H_{5'}H_{5''})^{-1},
g_{\tilde 2}=(h_{\tilde 5}H_5H_{5'})^{-1}$
&$g_{6}=(H_5H_{5'})^{-1}$
\\
&
&$g_{\tilde 3}=(h_{\tilde 5}H_5H_{5''})^{-1},
 g_{\tilde 4}=(h_{\tilde 5}H_{5'}H_{5''})^{-1}$
&$g_{7}=(H_5H_{5''})^{-1}$
\\
&
&$g_{\tilde 5}=h_{\tilde 5}^{-1}$
&$g_{8}=(H_{5'}H_{5''})^{-1}$
\\
\hline
(b)& 2
&$g_{\tilde 1}=(h_{\tilde 5}H_5H_{5'}H_{5''})^{-1}$
& $g_{4}=(H_{5}H_{5'}H_{5''})^{-1}$
\\
&
&
$g_{\tilde 2}=(h_{\tilde 5}H_5H_{5'})^{-1},
g_{\tilde 3}=(h_{\tilde 5}H_5H_{5''})^{-1}$,
&$g_{5}=H_{5}^{-1}$
\\
&
&$
g_{\tilde 6}=(h_{\tilde 5}H_{5'})^{-1},
g_{\tilde 7}=(h_{\tilde 5}H_{5''})^{-1}$
&$g_{8}=(H_{5'}H_{5''})^{-1}$
\\
\hline
(c)& 1
&$g_{\tilde 1}=(h_{\tilde 5}H_5H_{5'}H_{5''})^{-1},
g_{\tilde 2}=(h_{\tilde 5}H_5H_{5'})^{-1}$
&$g_{5}=(H_{5}H_{5'}H_{5''})^{-1}$
\\
  &
&$g_{\tilde 3}=(h_{\tilde 5}H_5H_{5''})^{-1},
 g_{\tilde 4}=(h_{\tilde 5}H_{5'}H_{5''})^{-1}$
&$g_{7}=H_{5}^{-1}, g_{8}=H_{5'}^{-1}$
\\
&
&$g_{\tilde 6}=h_{\tilde 5}^{-1}$
&$g_{9}=H_{5''}^{-1}$
\\
\hline
(d)& 1
&$g_{\tilde 1}=g_{\tilde 2}=(h_{\tilde 5}H_5H_{5'}H_{5''})^{-1}$
&$g_{6}=(H_5H_{5'}H_{5''})^{-1}$
\\
  &
&$g_{\tilde 3}=(h_{\tilde 5}H_5)^{-1},
 g_{\tilde 4}=(h_{\tilde 5}H_{5'})^{-1}$
&$g_{7}=H_{5}^{-1}, g_{8}=H_{5'}^{-1} $
\\
  &
&$g_{\tilde 5}=(h_{\tilde 5}H_{5''})^{-1}$
& $g_{9}=H_{5''}^{-1}$
\\
\hline
(e)& 1
&$g_{\tilde 1}=g_{\tilde 2}=(h_{\tilde 5}H_5H_{5'}H_{5''})^{-1}$
&$g_{4}=(H_5H_{5'})^{-1}$
\\
  &
&$g_{\tilde 3}=(h_{\tilde 5}H_5)^{-1},
 g_{\tilde 6}=(h_{\tilde 5}H_{5'}H_{5''})^{-1}$
&$g_{5}=(H_{5}H_{5''})^{-1} $
\\
  &
&$g_{\tilde 7}=(h_{\tilde 5})^{-1}$
& $g_{8}=H_{5'}^{-1}, g_{9}=H_{5''}^{-1}$
\\
\hline
(f)& 2
&$g_{\tilde 1}=g_{\tilde 2}=(h_{\tilde 5}H_5H_{5'}H_{5''})^{-1}$
&$g_{6}=(H_5 H_{5'})^{-1}$
\\
  &
&$g_{\tilde 3}=(h_{\tilde 5}H_5)^{-1},
 g_{\tilde 4}=(h_{\tilde 5}H_{5'})^{-1}$
&$g_{7}=(H_{5}H_{5''})^{-1}$
\\
  &
&$g_{\tilde 5}=(h_{\tilde 5}H_{5''})^{-1}$
&$g_{8}=(H_{5'}H_{5''})^{-1}$
\\
\hline
\hline
\multicolumn{2}{|c||}{M2M5$^3$}&
${\cal A}=h_{\tilde 2}^{1/3}(H_5H_{5'}H_{5''})^{2/3}$&
$g_{0}=(h_{\tilde 2}H_5H_{5'}H_{5''})^{-1}$
\\
\hline
&${\rm dim}(Z)$&
$g_{\tilde \alpha}$&$g_{\alpha}$
\\
\hline
(a)& 2
&$g_{\tilde 1}= (h_{\tilde 2}H_5H_{5'}H_{5''})^{-1}$
&$g_{3}=g_{4}=H_5^{-1}H_{5'}^{-1}$
\\
 &
&$g_{\tilde 2}=h_{\tilde 2}^{-1}$
&$g_{5}=g_{6}=H_5^{-1}H_{5''}^{-1}$
\\
&
&
&$g_{7}=g_{8}=H_{5'}^{-1}H_{5''}^{-1}$
\\
\hline
(c)& 1
&$g_{\tilde 1}= (h_{\tilde 2}H_5H_{5'}H_{5''})^{-1}$
&$g_{3}=(H_5H_{5'}H_{5''})^{-1},
g_{4}=H_5^{-1}H_{5'}^{-1}$
\\
 &
&$g_{\tilde 2}=h_{\tilde 2}^{-1}$
&$g_{5}=H_{5'}^{-1}H_{5''}^{-1},
g_{6}=H_5^{-1}H_{5''}^{-1}$
\\
&
&
&$g_{7}=H_{5}^{-1},
g_{8}=H_{5'}^{-1},
g_{9}=H_{5''}^{-1}$
\\
\hline
(e)& 2
&$g_{\tilde 1}=h_{\tilde 2}^{-1}H_{5}^{-1}H_{5'}^{-1}$
&$g_{3}=g_{4}=(H_5H_{5'}H_{5''})^{-1}$
\\
&
&$g_{\tilde 2}=h_{\tilde 2}^{-1}H_{5''}^{-1}$
&$g_{5}=H_5^{-1}H_{5''}^{-1}, g_{6}=H_{5'}^{-1}H_{5''}^{-1}$
\\
 &
&
&$g_{7}=H_{5}^{-1},
g_{8}=H_{5'}^{-1}$
\\
\hline
\hline
\multicolumn{2}{|c||}{M2M5$^3$}&
${\cal A}=h_{\tilde 5}^{2/3}H_2^{1/3}H_{5}^{2/3}H_{5'}^{2/3}$&
$g_{0}= (h_{\tilde 5}H_2H_{5}H_{5'})^{-1}$
\\
\hline
&${\rm dim}(Z)$&
$g_{\tilde \alpha}$&$g_{\alpha}$
 \\
\hline
(b)& 2
&$g_{\tilde 1}= (h_{\tilde 5}H_2H_{5}H_{5'})^{-1}$
&$g_{2}=H_2^{-1}$
\\
&
&$g_{\tilde 3}=g_{\tilde 4}=h_{\tilde 5}^{-1}H_{5}^{-1}$
&$g_{7}=g_{8}=H_{5}^{-1}H_{5'}^{-1}$
\\
&
&$g_{\tilde 5}=g_{\tilde 6}=h_{\tilde 5}^{-1}H_{5'}^{-1}$
&
\\
\hline
(d)& 1
&$g_{\tilde 1}=(h_{\tilde 5}H_2H_{5}H_{5'})^{-1}$
&$g_{2}=H_2^{-1}$
\\
 &
&$g_{\tilde 3}=h_{\tilde 5}^{-1}(H_{5}H_{5'})^{-1},
g_{\tilde 4}=h_{\tilde 5}^{-1} H_{5}^{-1}$
&$g_{5}=H_{5}^{-1}H_{5'}^{-1}$
\\
 &
&$g_{\tilde 6}=h_{\tilde 5}^{-1}H_{5'}^{-1},
g_{\tilde 7}=h_{\tilde 5}^{-1} $
&$g_{8}=H_{5}^{-1},
g_{9}=H_{5'}^{-1}$
\\
\hline
(f)& 2
&$g_{\tilde 1}=h_{\tilde 5}^{-1}H_2^{-1}H_{5}^{-1}$
&$g_{2}=H_2^{-1}H_{5'}^{-1}$
\\
 &
&$g_{\tilde 3}=g_{\tilde 4}=h_{\tilde 5}^{-1}H_5^{-1}H_{5'}^{-1}$
&$g_{6}=H_{5}^{-1}H_{5'}^{-1}$
\\
 &
&$g_{\tilde 5}=h_{\tilde 5}^{-1}H_{5'}^{-1},
g_{\tilde 7}=h_{\tilde 5}^{-1}$
&$g_{8}=H_{5}^{-1}$
\\
\hline
(g)& 2
&$g_{\tilde 2}=h_{\tilde 5}^{-1}H_2^{-1}$
&$g_{1}=H_2^{-1}H_{5}^{-1}H_{5'}^{-1}$
\\
 &
&$g_{\tilde 3}=g_{\tilde 4}=h_{\tilde 5}^{-1}H_5^{-1}H_{5'}^{-1}$
&$g_{7}=H_{5}^{-1}$
\\
 &
&$g_{\tilde 5}=h_{\tilde 5}^{-1}H_{5}^{-1},
g_{\tilde 6}=h_{\tilde 5}^{-1}H_{5'}^{-1}$
&$g_{8}=H_{5'}^{-1}$
\\
\hline
\end{tabular}
}
\label{table_4-12}
\end{center}
\end{table}


\begin{table}[ht]
\caption{Intersections of four M-branes II.}
{\scriptsize
\begin{center}
\begin{tabular}{|c||c|c|c|c|c|c|c|c|c|c|c|c||c||c|c|}
\hline
&&0&1&2&3&4&5&6&7&8&9&10&$\tilde{I}$&cos&BH
\\
\hline
&M2 & $\circ$ & $\circ$ &  &  &  &  &&&$\circ$&&& (a)
&$-$&$-$
\\
\cline{3-13}
&M2 & $\circ$ & & $\circ$ & &&&  &  & & $\circ$&&&&
\\
\cline{3-13}
&M5 & $\circ$ & $\circ$ & $\circ$  &$\circ$& $\circ$ &$\circ$&  &  & & &&(b)
&$-$&$-$
\\
\cline{3-13}
&M5 & $\circ$ &$\circ$&  $\circ$ & $\circ$ &&& $\circ$ & $\circ$& &&& &&
\\
\cline{2-16}
&M2 & $\circ$ & $\circ$ &  &   &   & & &&$\circ$&&& (c)
&$-$&$-$
\\
\cline{3-13}
&M2 & $\circ$ &   &   &  &$\circ$&
& $\circ$ &   & & &&(d)&$-$&$-$
\\
\cline{3-13}
(M2)$^2$
(M5)$^2$
&M5 & $\circ$ & $\circ$ & $\circ$  &$\circ$& $\circ$ & $\circ$&   & &   & &&&&
\\
\cline{3-13}
&M5 & $\circ$ &$\circ$& $\circ$  & $\circ$ & & & $\circ$ &$\circ$ &  & &&(e)
&$-$&$-$
\\
\cline{2-16}
&M2 & $\circ$ &  & &  & $\circ$ & &$\circ$&&&&&(f)
&$\surd$&$\surd$
\\
\cline{3-13}
&M2 & $\circ$ &   &   &   && $\circ$
&  & $\circ$ & & && &&
\\
\cline{3-13}
&M5 &  $\circ$ & $\circ$  &  $\circ$  &$\circ$& $\circ$ & $\circ$&  &
&  & &&(g)
&$\surd$&$\surd$
\\
\cline{3-13}
&M5 &  $\circ$ & $\circ$&  $\circ$  & $\circ$ && & $\circ$ &$\circ$ &  &&&&&
\\
\hline
\hline
&M2 & $\circ$ & $\circ$ & $\circ$ &  &  &   &&&&&&  (a)
&$-$&$-$
\\
\cline{3-13}
(M2)$^3$
M5
&M2 & $\circ$ &   &   & $\circ$ &$\circ$ && &   & & && &&
\\
\cline{3-13}
&M2 & $\circ$ &   &    &&   &$\circ$ & $\circ$ &   & & &&&&
\\
\cline{3-13}
&M5 & $\circ$ &$\circ$&   & $\circ$ & &$\circ$&  & $\circ$&  $\circ$& &&
(b)&$-$&$-$
\\
\hline
\hline
&M2 & $\circ$ & $\circ$ & $\circ$ &   &   & & &&&&& (a)
&$-$&$-$
\\
\cline{3-13}
&M2 & $\circ$ &   &   & $\circ$ &$\circ$&
&  &   & & &&&&
\\
\cline{3-13}
(M2)$^4$
&M2 & $\circ$ &   &    &&   & $\circ$& $\circ$ & &  & &&&&
\\
\cline{3-13}
&M2 & $\circ$ & &   &   & & &  &$\circ$ & $\circ$ & &&&&
\\
\hline
\end{tabular}
\label{table_4-2}
\end{center}
}
\end{table}
\setcounter{table}{10}
\begin{table}[ht]
\caption{(Continue) Concrete metrics for four M-branes II.}
\begin{center}
{\scriptsize
\begin{tabular}{|c|c||c|c|}
\hline
\multicolumn{2}{|c||}{M2$^2$M5$^2$}&
${\cal A}=(h_{\tilde 2}H_2)^{1/3}(H_{5}H_{5'})^{2/3}$&
$g_{\tilde 0}=(h_{\tilde 2}H_2H_{5}H_{5'})^{-1}$
\\
\hline
&${\rm dim}(Z)$&
$g_{\tilde \alpha}$&$g_{\alpha}$
\\
\hline
(a)& 1
&$g_{\tilde 1}=(h_{\tilde 2}H_5H_{5'})^{-1}$
&$g_{2}=(H_{2}H_5H_{5'})^{-1},g_{3}=(H_5H_{5'})^{-1},
$
\\
  &
&$g_{8}=h_{\tilde 2}^{-1}$
&$g_{4}=g_{5}=H_5^{-1}, g_{6}=g_{7}=H_{5'}^{-1}$
\\
  &
&
&$g_{9}=H_{2}^{-1}$
\\
\hline
(c)& 2
&$g_{\tilde 1}=(h_{\tilde 2}H_5H_{5'})^{-1}$
& $g_{2}=
g_{3}=(H_5H_{5'})^{-1}$
\\
&
&$g_{8}=h_{\tilde 2}^{-1}$
&$g_{4}=(H_{2}H_5)^{-1},
g_{5}=H_5^{-1}$
\\
&
&
&$g_{6}=(H_{2}H_{5'})^{-1},
g_{7}=H_{5'}^{-1}$
\\
\hline
(d)& 2
&$g_{\tilde 4}=(h_{\tilde 2}H_5H)^{-1}$
& $g_{1}=(H_{2}H_5H_{5'})^{-1}$
\\
&
&$g_{\tilde 6}=(h_{\tilde 2}H_{5'})^{-1}$,
&$g_{2}=g_{3}=(H_5H_{5'})^{-1}$
\\
&
&
&$g_{5}=H_5^{-1}, g_{7}=H_{5'}^{-1},
g_{8}=H_{2}^{-1}$
\\
\hline
(f)& 3
&$g_{\tilde 4}=(h_{\tilde 2}H_5)^{-1}$
& $g_{1}=g_{2}=g_{3}=(H_5H_{5'})^{-1}$
\\
&
&$g_{\tilde 6}=(h_{\tilde 2}H_{5'})^{-1}$,
&$g_{5}=(H_2H_{5})^{-1},
g_{7}=(H_2H_{5'})^{-1}$
\\
\hline
\hline
\multicolumn{2}{|c||}{M2$^2$M5$^2$}&
${\cal A}=h_{\tilde 5}^{2/3}(H_2H_{2'})^{1/3}H_{5}^{2/3}$&
$g_{0}= (h_{\tilde 5}H_2H_{2'}H_{5})^{-1}$
\\
\hline
&${\rm dim}(Z)$&
$g_{\tilde \alpha}$&$g_{\alpha}$
 \\
\hline
(b)& 1
&$g_{\tilde 1}= (h_{\tilde 5}H_2H_{5})^{-1}$
&$g_{6}=g_{7}=H_{5}^{-1}$
\\
&
&$ g_{\tilde 2}=(h_{\tilde 5}H_{2'}H_{5})^{-1}$
&$g_{8}=H_2^{-1}$
\\
&
&$g_{\tilde 3}=(h_{\tilde 5}H_{5})^{-1},
g_{\tilde 4}=g_{\tilde 5}=h_{\tilde 5}^{-1}$
&$g_{9}=H_{2'}^{-1}$
\\
\hline
(e)& 2
&$g_{\tilde 1}= (h_{\tilde 5}H_2H_{5})^{-1}$
&$g_{6}=(H_2H_{5})^{-1}$
\\
&
&$ g_{\tilde 2}=g_{\tilde 3}=(h_{\tilde 5}H_{5})^{-1}$
&$g_{7}=H_5^{-1}$
\\
&
&$ g_{\tilde 4}=(h_{\tilde 5}H_{2'})^{-1},
g_{\tilde 5}=h_{\tilde 5}^{-1}$
&$g_{8}=H_{2}^{-1}$
\\
\hline
(g)& 3
&$g_{\tilde 1}= g_{\tilde 2}=g_{\tilde 3}=(h_{\tilde 5}H_{5})^{-1}$
&$g_{6}=(H_{2}H_{5})^{-1}$
\\
&
&$ g_{\tilde 4}=(h_{\tilde 5}H_{2})^{-1},
g_{\tilde 5}=(h_{\tilde 5}H_{2'})^{-1}$
&$g_{7}=(H_{2'}H_5)^{-1}$
\\
\hline
\hline
\multicolumn{2}{|c||}{M2$^3$M5}&
${\cal A}=h_{\tilde 2}^{1/3}(H_{2}H_{2'}H_5)^{2/3}$&
$g_{0}=(h_{\tilde 2}H_{2}H_{2'}H_5)^{-1}$
\\
\hline
&${\rm dim}(Z)$&
$g_{\tilde \alpha}$&$g_{\alpha}$
\\
\hline
(a)& 2
&$g_{\tilde 1}= (h_{\tilde 2}H_5)^{-1}$
&$g_{3}=(H_{2}H_5)^{-1},
g_{4}=H_{2}^{-1}$
\\
 &
&$g_{\tilde 2}=h_{\tilde 2}^{-1}$
&$g_{5}=(H_{2'}H_5)^{-1},
g_{6}=H_{2'}^{-1}$
\\
&
&
&$g_{7}=g_{8}=H_5^{-1}$
\\
\hline
\hline
\multicolumn{2}{|c||}{M2$^3$M5}&
${\cal A}=h_{\tilde 5}^{2/3}(H_2H_{2'}H_{2''})^{1/3}$&
$g_{0}= (h_{\tilde 5}H_2H_{2'}H_{2''})^{-1}$
\\
\hline
&${\rm dim}(Z)$&
$g_{\tilde \alpha}$&$g_{\alpha}$
 \\
\hline
(b)& 2
&$g_{\tilde 1}=(h_{\tilde 5}H_2)^{-1},
g_{\tilde 3}=(h_{\tilde 5}H_{2'})^{-1}$
&$g_{2}=H_2^{-1}$
\\
&
&$g_{\tilde 5}=(h_{\tilde 5}H_{2''})^{-1}$
&$g_{4}=H_{2'}^{-1}$
\\
&
&$
g_{\tilde 7}=g_{\tilde 8}=h_{\tilde 5}^{-1}$
&$g_{6}=H_{2''}^{-1}$
\\
\hline
\hline
\multicolumn{2}{|c||}{M2$^4$}&
${\cal A}=(h_{\tilde 2}H_2H_{2'}H_{2''})^{2/3}$&
$g_{0}= (h_{\tilde 2}H_2H_{2'}H_{2''})^{-1}$
\\
\hline
&${\rm dim}(Z)$&
$g_{\tilde \alpha}$&$g_{\alpha}$
 \\
\hline
(a)& 2
&
&$g_{3}= g_{4}=H_{2}^{-1}$
\\
&
&$g_{\tilde 1}= g_{\tilde 2}=h_{\tilde 2}^{-1}$
&$g_{5}= g_{6}=H_{2'}^{-1}$
\\
&
&
&$g_{7}= g_{8}=H_{2''}^{-1}$
\\
\hline
\end{tabular}
}
\end{center}
\end{table}


\begin{table}[ht]
\caption{
Intersections of five M-branes.
}
\begin{center}
{\scriptsize
\begin{tabular}{|c||c|c|c|c|c|c||c|c|c|}
\hline
branes&dim($\Zsp$)&1M&2M&3M&4M&5M
&$\tilde I(\#)$&cos&BH\\
\hline
\hline
&1&2&2&2&2&1 & M5(3)& $-$  &$-$ \\
\cline{2-10}
(M5)$^5$&1&1&4&2&0&2& M5(2)& $-$ &$-$\\
\cline{2-10}
&2&0&2&4&1&1&M5(2)& $-$  &$-$ \\
\hline
&\raisebox{-.5em}{1}&\raisebox{-.5em}{3}&\raisebox{-.5em}{1}
&\raisebox{-.5em}{3}&\raisebox{-.5em}{2}&\raisebox{-.5em}{0}
&M2(1)& $-$ &$-$\\[-.2em]
\cline{8-10}
&&&&&&&M5(2)&$-$  &$-$\\
\cline{2-10}
&\raisebox{-.5em}{2}&\raisebox{-.5em}{0}&\raisebox{-.5em}{4}
&\raisebox{-.5em}{2}&\raisebox{-.5em}{2}&\raisebox{-.5em}{0}
&M2(1)&$-$  &$-$\\[-.2em]
\cline{8-10}
\raisebox{-.5em}{M2(M5)$^4$}
&&&& & &&M5(1)&$-$  &$-$\\[-.4em]
\cline{2-10}
&\raisebox{-.5em}{1}&\raisebox{-.5em}{1}&\raisebox{-.5em}{6}
&\raisebox{-.5em}{0}&\raisebox{-.5em}{1}&\raisebox{-.5em}{1}
&M2(1)& $-$ &$-$\\[-.2em]
\cline{8-10}
&&&& &&&M5(1) &$-$  &$-$\\
\cline{2-10}
&\raisebox{-.5em}{1}&\raisebox{-.5em}{2}&\raisebox{-.5em}{3}
&\raisebox{-.5em}{3}&\raisebox{-.5em}{0}
&\raisebox{-.5em}{1}&M2(1)& $-$ &$-$\\[-.2em]
\cline{8-10}
&&&&&&&M5(2)& $-$ &$-$\\
\hline
&\raisebox{-.5em}{1}&\raisebox{-.5em}{3}&\raisebox{-.5em}{3}
&\raisebox{-.5em}{2}&\raisebox{-.5em}{1}&\raisebox{-.5em}{0}
&M2(2)& $-$ &\\[-.2em]
\cline{8-10}
\raisebox{-.5em}{(M2)$^2$(M5)$^3$}
&&&&&&&M5(2)& $-$ &$-$\\[-.4em]
\cline{2-10}
&\raisebox{-.5em}{2}&\raisebox{-.5em}{1}&\raisebox{-.5em}{3}
&\raisebox{-.5em}{4}&\raisebox{-.5em}{0}&\raisebox{-.5em}{0}
&M2(1)& $-$ &$-$\\[-.2em]
\cline{8-10}
&&&&&&&M5(2)&$-$  &$-$\\
\hline
&\raisebox{-.5em}{1}&\raisebox{-.5em}{4}&\raisebox{-.5em}{3}
&\raisebox{-.5em}{2}&\raisebox{-.5em}{0}&\raisebox{-.5em}{0}
&M2(2)& $-$ &$-$\\[-.2em]
\cline{8-10}
\raisebox{-.5em}{(M2)$^3$(M5)$^2$}
&&&&&&&M5(1)& $-$ &$-$\\[-.4em]
\cline{2-10}
&\raisebox{-.5em}{2}&\raisebox{-.5em}{1}&\raisebox{-.5em}{6}
&\raisebox{-.5em}{1}&\raisebox{-.5em}{0}&\raisebox{-.5em}{0}
&M2(2)&$-$  &$-$\\[-.2em]
\cline{8-10}
&&&&&&&M5(1)&  $-$  &$-$  \\
\hline
\raisebox{-.5em}{(M2)$^4$(M5)$^1$}&\raisebox{-.5em}{1}
&\raisebox{-.5em}{5}&\raisebox{-.5em}{4}&\raisebox{-.5em}{0}
&\raisebox{-.5em}{0}&\raisebox{-.5em}{0}&M2(1)&$-$ &$-$\\[-.4em]
\cline{8-10}
&&&&&&&M5(1)& $-$ &$-$\\
\hline
\end{tabular}
}
\label{table_5}
\end{center}
\end{table}
\begin{table}[ht]
\caption{
Intersections of six M-branes.
}
\begin{center}
{\scriptsize
\begin{tabular}{|c||c|c|c|c|c|c|c||c|c|c|}
\hline
branes&dim($\Zsp$)&1M&2M&3M&4M&5M&6M
&$\tilde I(\#)$&cos&BH\\
\hline
&1&0&3&4&0&0&2&M5(1) & $-$&$-$  \\
\cline{2-11}
(M5)$^6$&1&1&2&2&2&1&1 &
M5(3)&  $-$ & $-$
\\
\cline{2-11}
&2&0&0&4&3&0&1&
M5(1)& $-$ & $-$
\\
\hline
\raisebox{-.5em}{M2(M5)$^5$}
&\raisebox{-.5em}{1}&\raisebox{-.5em}{1}&\raisebox{-.5em}{2}
&\raisebox{-.5em}{4}&\raisebox{-.5em}{1}&\raisebox{-.5em}{0}
&\raisebox{-.5em}{1}&
M2(1)&  $-$  & $-$
\\[-.2em]
\cline{9-11}
&&&&&&&&
M5(2)&  $-$  & $-$
\\
\hline
&\raisebox{-.5em}{1}&\raisebox{-.5em}{1}&\raisebox{-.5em}{4}
&\raisebox{-.5em}{2}&\raisebox{-.5em}{1}& \raisebox{-.5em}{1}
& \raisebox{-.5em}{0}&
M2(2)& $-$  & $-$\\[-.2em]
\cline{9-11}
&&&&&&&&
M5(1)& $-$ & $-$\\
\cline{2-11}
\raisebox{-.5em}{(M2)$^2$(M5)$^4$}
&\raisebox{-.5em}{1}&\raisebox{-.5em}{2}&\raisebox{-.5em}{2}
&\raisebox{-.5em}{2}&\raisebox{-.5em}{3}&\raisebox{-.5em}{0}
&\raisebox{-.5em}{0}&
M2(1)&  $-$  & $-$ \\[-.4em]
\cline{9-11}
&&&&&&& &
M5(2)& $-$  & $-$ \\
\cline{2-11}
&\raisebox{-.5em}{2}&\raisebox{-.5em}{0}&\raisebox{-.5em}{2}
&\raisebox{-.5em}{4}&\raisebox{-.5em}{2}&\raisebox{-.5em}{0}
&\raisebox{-.5em}{0}&
M2(1)&  $-$  & $-$
\\[-.2em]
\cline{9-11}
&&&&&&&&
M5(1)& $-$  & $-$
\\
\hline
&\raisebox{-.5em}{1}&\raisebox{-.5em}{2}&\raisebox{-.5em}{3}
&\raisebox{-.5em}{3}&\raisebox{-.5em}{1}&\raisebox{-.5em}{0}&
\raisebox{-.5em}{0}&
M2(2)& $-$   & $-$ \\[-.2em]
\cline{9-11}
(M2)$^3$(M5)$^3$&&&&&&&&
M5(2)& $-$  & $-$ \\
\cline{2-11}
&\raisebox{-.5em}{2}&\raisebox{-.5em}{0}&\raisebox{-.5em}{3}
&\raisebox{-.5em}{5}&\raisebox{-.5em}{0}&\raisebox{-.5em}{0}
&\raisebox{-.5em}{0}&
M2(1)&  $-$  & $-$
\\[-.2em]
\cline{9-11}
&&&&&&&&
M5(1)& $-$  & $-$\\
\hline
\raisebox{-.5em}{(M2)$^4$(M5)$^2$}&
\raisebox{-.5em}{1}&\raisebox{-.5em}{2}&\raisebox{-.5em}{5}
&\raisebox{-.5em}{2}&\raisebox{-.5em}{0}&\raisebox{-.5em}{0}
&\raisebox{-.5em}{0}&
M2(2)&  $-$  & $-$ \\[-.4em]
\cline{9-11}
&&&&&&&&
M5(1)& $-$  & $-$
\\
\hline
\end{tabular}
}
\label{table_6}
\end{center}
\end{table}
\begin{table}[ht]
\caption{
Intersections of seven M-branes.
}
\begin{center}
{\scriptsize
\begin{tabular}{|c||c|c|c|c|c|c|c|c||c|c|c|}
\hline
branes&dim($\Zsp$)&1M&2M&3M&4M&5M&6M&7M
&$\tilde I(\#)$&cos&BH\\
\hline
&1&1&0&4&0&3&0&1 &M5(2)&$-$&$-$ \\
\cline{2-12}
\raisebox{-.5em}{(M5)$^7$}&1&0&3&0&4&0&1&1 &M5(2)&$-$&$-$\\[-.4em]
\cline{2-12}
&1&0&0&7&0&0&0&2 &M5(2)&$-$&$-$ \\
\cline{2-12}
&2&0&0&0& 7& 0& 0&1 &M5(2)&$-$&$-$ \\
\hline
\raisebox{-.5em}{M2(M5)$^6$}&\raisebox{-.5em}{1}
&\raisebox{-.5em}{1}&\raisebox{-.5em}{0}& \raisebox{-.5em}{4}
& \raisebox{-.5em}{3}& \raisebox{-.5em}{0}& \raisebox{-.5em}{0}
& \raisebox{-.5em}{1} &
M2(1)&$-$&$-$ \\[-.4em]
\cline{10-12}
&&&&& & & &  & M5(1)&$-$&$-$ \\
\hline
&\raisebox{-.5em}{1}&\raisebox{-.5em}{1}& \raisebox{-.5em}{2}
& \raisebox{-.5em}{4}& \raisebox{-.5em}{1}& \raisebox{-.5em}{1}
& \raisebox{-.5em}{0}& \raisebox{-.5em}{0} &
M2(2)&$-$&$-$ \\[-.2em]
\cline{10-12}
&&&&&&&&&
M5(1)&$-$&$-$ \\
\cline{2-12}
\raisebox{-.5em}{(M2)$^3$(M5)$^4$}&\raisebox{-.5em}{1}
&\raisebox{-.5em}{1}& \raisebox{-.5em}{3}& \raisebox{-.5em}{1}
& \raisebox{-.5em}{4}& \raisebox{-.5em}{0}& \raisebox{-.5em}{0}
& \raisebox{-.5em}{0} &
M2(1)&$-$&$-$ \\[-.4em]
\cline{10-12}
&&&&&&&&&
M5(2)&$-$&$-$ \\
\cline{2-12}
&\raisebox{-.5em}{2}&\raisebox{-.5em}{0}& \raisebox{-.5em}{0}
& \raisebox{-.5em}{6}& \raisebox{-.5em}{2}&\raisebox{-.5em}{0}
& \raisebox{-.5em}{0}& \raisebox{-.5em}{0} &
M2(1)&$-$&$-$  \\[-.2em]
\cline{10-12}
&&&&&&&&&
M5(1)&$-$&$-$  \\
\hline
\raisebox{-.5em}{(M2)$^4$(M5)$^3$}&\raisebox{-.5em}{1}
&\raisebox{-.5em}{1}&\raisebox{-.5em}{3}& \raisebox{-.5em}{4}
& \raisebox{-.5em}{1}& \raisebox{-.5em}{0}& \raisebox{-.5em}{0}&
\raisebox{-.5em}{0} &
M2(2)&$-$&$-$ \\[-.4em]
\cline{10-12}
&&&&&&&&&
M5(1)&$-$&$-$ \\
\hline
\end{tabular}
}
\label{table_7}
\end{center}
\end{table}

\begin{table}[ht]
\caption{
Intersections of eight M-branes.
}
\begin{center}
{\scriptsize
\begin{tabular}{|c||c|c|c|c|c|c|c|c|c||c|c|c|}
\hline
branes&dim($\Zsp$)&1M&2M&3M&4M&5M&6M&7M&8M
&$\tilde I(\#)$&cos&BH\\
\hline
\raisebox{-.5em}{M2(M5)$^7$}
&\raisebox{-.5em}{1}&\raisebox{-.5em}{1}&\raisebox{-.5em}{0}
&\raisebox{-.5em}{0}&\raisebox{-.5em}{7}&\raisebox{-.5em}{0}
&\raisebox{-.5em}{0}&\raisebox{-.5em}{0}&\raisebox{-.5em}{1} &
M2(1)& $-$    &  $-$  \\[-.2em]
\cline{11-13}
&&&&&&&&&&
M5(1)&  $-$    &  $-$  \\
\hline
&\raisebox{-.5em}{1}&\raisebox{-.5em}{0}& \raisebox{-.5em}{4}
& \raisebox{-.5em}{0}& \raisebox{-.5em}{5}& \raisebox{-.5em}{0}
& \raisebox{-.5em}{0}& \raisebox{-.5em}{0}& \raisebox{-.5em}{0} &
M2(1)&  $-$  &  $-$  \\[-.2em]
\cline{11-13}
\raisebox{-.5em}{(M2)$^4$(M5)$^4$}&&&&&&&&&&
M5(1)&  $-$  &  $-$  \\[-0.4em]
\cline{2-13}
&\raisebox{-.5em}{1}&\raisebox{-.5em}{1}&\raisebox{-.5em}{0}
& \raisebox{-.5em}{6}& \raisebox{-.5em}{1}&\raisebox{-.5em}{1}
& \raisebox{-.5em}{0}& \raisebox{-.5em}{0}& \raisebox{-.5em}{0} &
M2(2)&  $-$ &  $-$  \\[-.2em]
\cline{11-13}
&&&&&&&&&&
M5(1)&  $-$ &  $-$  \\
\hline
\end{tabular}
}
\label{table_8}
\end{center}
\end{table}

\newpage

\section{Dynamical solution of KK-wave and KK-monopole}
\label{KK_and_monopole}

\subsection{KK-wave}
\label{Appendix:KKW}

Here, we discuss the dynamical solution of of KK-wave.
We start from $(D-1)$-dimensional spacetime,
and consider  the KK 2-form ${\cal F}_{AB}$ with a coupling to the dilaton.
Replacing $D$ with $(D-1)$ and
substituting $p=0$ into Eqs.
(\ref{gs:Einstein equations:Eq}), (\ref{gs:scalar field equation:Eq}),
(\ref{gs:gauge field equation:Eq}), (\ref{gs:paremeter:Eq}), and
(\ref{p:metric of p-brane solution:Eq}),
we find the electric 0-brane solution in $(D-1)$ dimensions  written by
\Eqrsubl{0b}{
&&
ds_{D-1}^2=-h_{\rm w}^{~-{D-4\over D-3}}dt^2+h_{\rm w}^{~{1\over D-3}}
\,u_{ij}(\Zsp) \,dz^idz^j,\\
&&
e^{\phi}=h_{\rm w}^{~\sqrt{D-2\over 2(D-3)}},~~~~~
{\cal A}^{\rm (w)}=(h_{\rm w}^{-1}-1) \, dt, \\
&&
R_{ij}(\Zsp)=0,
\\
&&
h_{\rm w}(t, z)=K_{\rm w}(t)+H_{\rm w}(z),
~~~~K_{\rm w}(t)=A_{\rm w} t+B_{\rm w},
~~~~\lap_{\Zsp}H_{\rm w}=0,
   \label{KKW:Dmetric:Eq}
}
where ${\cal F}=d{\cal A}^{\rm (w)}$,
and $R_{ij}(\Zsp)$ and $\lap_{\Zsp}$ are the Ricci tensor,
and the Laplace operator with respect to the $(D-2)$-dimensional
metric $u_{ij}$,
and $A_{\rm w}$ and  $B_{\rm w}$ are integration constants.
Before going to $D$ dimensions, we have to rescale the metric
\Eqref{KKW:Dmetric:Eq} to put it in the $D$-dimensional Einstein frame.
This is given by the conformal transformation
\Eq{
\bar{g}_{MN}=h_{\rm w}^{-1/(D-3)}g_{MN}.
}
Gathering the above results, we find the $D$-dimensional metric
of KK-wave~\cite{Brinkmann:1923}:
\begin{eqnarray}
ds^2=g_{MN}dx^Mdx^N
&=&-h_{\rm w}^{-1}dt^2
+h_{\rm w}\left[d\zeta +(h_{\rm w}^{-1}-1)dt\right]^2
+u_{ij}dz^idz^j
\nn
&=&-dt^2+d\zeta^2
+f_{\rm w}\left(dt-d\zeta\right)^2
+u_{ij}dz^idz^j
\,,
\label{KKW:metric:Eq}
\end{eqnarray}
where $u_{ij}$ denotes the $(D-2)$-dimensional metric depending
only on the transverse coordinate $z^i$,
while the function $h_{\rm w}\equiv 1+f_{\rm w}$
can depend on both $t$ and $z^i$.

Now we discuss the dynamical solution of Einstein equations
for $D$-dimensional metric (\ref{KKW:metric:Eq}).
Substituting the metric (\ref{KKW:metric:Eq})
into then vacuum Einstein equations, we obtain
\Eqrsubl{KKW:EE:Eq}{
&&-\lap_{\Zsp}h_{\rm w}+(h_{\rm w}-2)\pd_t^2 h_{\rm w}=0,\label{KKW:EEtt:Eq}\\
&&(1-h_{\rm w})\pd_t^2 h_{\rm w}+\lap_{\Zsp}h_{\rm w}=0,\label{KKW:EEtz:Eq}\\
&&\pd_t\pd_i h_{\rm w}=0,\label{KKW:EEta:Eq}\\
&&h_{\rm w}\pd_t^2 h_{\rm w}-\lap_{\Zsp} h_{\rm w}=0,\label{KKW:EEzz:Eq}\\
&&R_{ij}(\Zsp)=0,\label{KKW:EEij:Eq}
}
where $\lap_{\Zsp}$ and $R_{ij}(\Zsp)$ are  the Laplace operator and
the Ricci tensor with respect to the metric $u_{ij}$, respectively.
{}From \Eqref{KKW:EEta:Eq}, we get
\Eq{
h_{\rm w}(t, z)=K_{\rm w}(t)+H_{\rm w}(z).
  \label{KKW:h:Eq}
}
Eqs. (\ref{KKW:EEtt:Eq}), (\ref{KKW:EEtz:Eq}),
and (\ref{KKW:EEzz:Eq}) are written by linear combinations of terms
depending on both $t$ and $z^i$, and those depending only on $z^i$.
Then, in order to satisfy Eqs.~(\ref{KKW:EEtt:Eq}), (\ref{KKW:EEtz:Eq}),
and (\ref{KKW:EEzz:Eq}), we obtain
\Eq{
\pd_t^2 K_{\rm w}=0,~~~~\lap_{\Zsp} H_{\rm w}=0.
}
We note that the function $K_{\rm w}(t)$ depends on the linear function
of the time $t$.
The Einstein equations with the metric (\ref{KKW:metric:Eq})
are reduced to
\bea
&&h_{\rm w}(t, z)=K_{\rm w}(t)+H_{\rm w}(z),\nn
&&\pd_t^2 K_{\rm w}=0,\nn
&&\lap_{\Zsp}H_{\rm w}=0,\nn
&&R_{ij}({\Zsp})=0.
\ena
Especially, for the case of  $u_{ij}=\delta_{ij}$, we find  the solution
of the KK wave as
\begin{eqnarray}
&&ds^2=-dt^2+d\zeta^2
+f_{\rm w}\left(dt-d\zeta\right)^2
+u_{ij}dz^idz^j
\nn
&&
f_{\rm w}(t, z)\equiv h_{\rm w}(t, z)-1,
 \label{KKW:PS:Eq}
\end{eqnarray}
with
\begin{eqnarray}
&&h_{\rm w}(t, z)=K_{\rm w}(t)+H_{\rm w}(z),\nn
&&K_{\rm w}(t)=A_{\rm w} t+B_{\rm w}\\
&&H_{\rm w}(z)=C_{\rm w}+\sum_k\frac{Q_{{\rm w},k}}{|\bm{z}-\bm{z}_k|^{D-4}},
\nonumber
\end{eqnarray}
where $A_{\rm w}$, $B_{\rm w}$, $C_{\rm w}$, $Q_{\rm w}$ are constant
parameters and $\bm{z}_{k}$ represent the positions of the branes in Z space.

\subsection{KK-monopole}

Next we discuss the dynamical solution of KK-monopole~\cite{Sorkin:1983ns, Gross:1983hb}.
In the reduced $(D-1)$-dimensional picture, it has to be a
magnetically charged $(D-5)$-brane with a 2-form $F_{(D-3)}$.
Replacing $D$ with $(D-1)$ and substituting $p=(D-5)$
into Eqs. (\ref{gs:Einstein equations:Eq}),
(\ref{gs:scalar field equation:Eq}),
(\ref{gs:gauge field equation:Eq}), (\ref{gs:paremeter:Eq}), and
(\ref{p:metric of p-brane solution:Eq}),
the electric $(D-5)$-brane solution in $(D-1)$ dimensions can be written as
\bea
&&ds_{(D-1)}^2=h_{\rm m}^{-{1\over D-3}}q_{\mu\nu}dx^{\mu}dx^{\nu}
+h_{\rm m}^{D-4\over D-3}u_{ij}(\Zsp)dz^idz^j,\nn
&&e^{\phi}=h_{\rm m}^{~\sqrt{(D-2)\over 2(D-3)}},~~~~~
F_{(D-3)}=d(h_{\rm m}^{-1})\wedge \sqrt{-q}dx^0\wedge\cdots\wedge
dx^{D-5},\nn
&&R_{\mu\nu}(\Xsp)=0,~~~~R_{ij}(\Zsp)=0,\nn
&&h_{\rm m}(x, z)=K_{\rm m}(x)+H_{\rm m}(z),~~~~D_{\mu}D_{\nu}K_{\rm m}=0,
~~~~\lap_{\Zsp}H_{\rm m}=0,
   \label{KKM:Dmetric:Eq}
\ena
where $R_{\mu\nu}(\Xsp)$, $D_{\mu}$ and $q$ are Ricci tensor, covariant
derivative, determinant constructed from the $(D-4)$-dimensional
metric $q_{\mu\nu}$ which depends only on the coordinate $x^{\mu}$,
$R_{ij}(\Zsp)$ and $\lap_{\Zsp}$ are Ricci tensor,
Laplace operator with respect to the three-dimensional
metric $u_{ij}$ which depends only on the coordinate $z^i$.
Before going to $D$ dimensions, we have to rescale the metric
\Eqref{KKM:Dmetric:Eq} to put it in the $D$-dimensional Einstein frame.
Then we use the conformal transformation
\Eq{
\bar{g}_{MN}=h_{\rm m}^{-1/(D-3)}g_{MN}.
}
Collecting the above results, we find the $D$-dimensional metric
of KK-monopole:
\bea
ds^2&=&g_{MN}dx^Mdx^N\nn
&=&q_{\mu\nu}(\Xsp)dx^{\mu}dx^{\nu}
+h_{\rm m}^{-1}\left(d\zeta+{\cal A}^{\rm (m)}_idz^i\right)^2
+h_{\rm m} u_{ij}(\Zsp)dz^idz^j
\,,
\label{KKM:metric:Eq}
\ena
where $q_{\mu\nu}$ is $(D-4)$-dimensional metric depends
only on the coordinate $x^{\mu}$, and $u_{ij}$ denotes the
three-dimensional metric depends only on the transverse
coordinate $z^i$, and the function $h_{\rm m}$
depends on $x^{\mu}$ as well as
$z^i$, and relation between $h_{\rm m}$ and ${\cal A}^{\rm (m)}_i$ is
\Eqr{
{\cal F}_{ij}\equiv \pd_i{\cal A}^{\rm (m)}_j
-\pd_j{\cal A}^{\rm (m)}_i=-\epsilon_{ijk}\, \pd^kh_{\rm m}.
  \label{KKM:potential:Eq}
}
Substituting the metric \Eqref{KKM:metric:Eq} into the $D$-dimensional
vacuum Einstein equations, we obtain
\Eqrsubl{KKM:EE:Eq}{
&&R_{\mu\nu}(\Xsp)-h_{\rm m}^{-1}D_{\mu}D_{\nu}h_{\rm m}=0,
\label{KKM:EEmn:Eq}\\
&&h_{\rm m}^{-1}\pd_{\mu}\pd_i h_{\rm m}=0,\label{KKM:EEmi:Eq}\\
&&h_{\rm m}^{-2}\lap_{\Xsp} h_{\rm m}=0,\label{KKM:EEzz:Eq}\\
&&{\cal A}^{\rm (m)}_ih_{\rm m}^{-2}\lap_{\Xsp}h
+{\cal A}^{\rm (m)}_ih_{\rm m}^{-3}\lap_{\Zsp}h_{\rm m}=0,
\label{KKM:EEzi:Eq}\\
&&R_{ij}(\Zsp)-\frac{1}{2}\left(u_{ij}-h_{\rm m}^{-2}{\cal A}^{\rm (m)}_i
{\cal A}^{\rm (m)}_j\right)\lap_{\Xsp}h_{\rm m}
-\frac{1}{2}h_{\rm m}^{-1}u_{ij}\lap_{\Zsp}h_{\rm m}=0,
\label{KKM:EEij:Eq}
}
where $D_{\mu}$, $\lap_{\Xsp}$, $R_{\mu\nu}(\Xsp)$ are covariant derivative,
Laplace operator, Ricci tensor with respect to the metric
$q_{\mu\nu}$, and $\lap_{\Zsp}$, $R_{ij}(\Zsp)$ are Laplace operator,
Ricci tensor with respect to the metric $u_{ij}$, and we assume
\Eq{
\pd_{\mu}{\cal A}^{\rm (m)}_i=0.
\label{KKM:A:Eq}
}
Using Eq.~(\ref{KKM:EEmi:Eq}), we get
\Eq{
h_{\rm m}(x, z)=K_{\rm m}(x)+H_{\rm m}(z).
  \label{KKM:h:Eq}
}
Equations \Eqref{KKM:EEmn:Eq}, \Eqref{KKM:EEzi:Eq},
\Eqref{KKM:EEij:Eq} are written by the combination of the term
depending not only on $x^{\alpha}$ but $z^i$, and depending only on $z^i$.
Then, in order to satisfy Eqs.~(\ref{KKM:EEmn:Eq}),
(\ref{KKM:EEmi:Eq}), and  (\ref{KKM:EEzz:Eq}), we choose
\Eq{
D_{\mu}D_{\nu}K=0,~~~~\lap_{\Zsp}H=0,~~~~R_{\alpha\beta}(\Xsp)=0,
~~~~R_{ij}({\Zsp})=0.
}
The Einstein equations in the metric \Eqref{KKM:metric:Eq}
are then reduced to
\bea
&&h_{\rm m}(x, z)=K_{\rm m}(x)+H_{\rm m}(z),\nn
&&D_{\mu}D_{\nu}K_{\rm m}=0,\nn
&&\lap_{\Zsp}H_{\rm m}=0,\nn
&&R_{\mu\nu}(\Xsp)=0,~~~~R_{ij}({\Zsp})=0.
 \label{KKM:Einstein:Eq}
\ena
For $q_{\mu\nu}=\eta_{\mu\nu}$, $u_{ij}=\delta_{ij}$,
we can obtain the solution of Einstein equation explicitly
\bea
&&ds^2=\eta_{\mu\nu}dx^{\mu}dx^{\nu}
+h_{\rm m}(dz^i)^2+h_{\rm m}^{-1}(d\zeta+{\cal A}^{\rm (m)}_idz^i)^2,
 \label{KKM:PS:Eq}
\ena
with
\bea
\hspace{-1cm}
\hspace{-1cm}&&h_{\rm m}(x, z)=K_{\rm m}(x)+H_{\rm m}(z),\nn
\hspace{-1cm}&&K_{\rm m}(x)=A_{{\rm m}\,(\mu)}\,x^{\mu}+B_{\rm m},\nn
\hspace{-1cm}&&H_{\rm m}(z)=C_{\rm m}+\sum_k
\frac{Q_{{\rm m},\, k}}{|\bm{z}-\bm{z}_k|},
\ena
where $A_{{\rm m}\,(\mu)}$, $B_{\rm m}$, $C_{\rm m}$, $Q_{{\rm m},\, k}$,
$\bm{z}_k$'s are integration constants.

\section{Intersecting branes with M-waves and KK-monopoles}
\label{Intersecting_branes_with_MKK}

A complete list for static brane system with M-waves and KK-monopoles
are given in~\cite{Bergshoeff:1997tt}.
Hence we pick up only interesting cases in which one can discuss
cosmology or a black hole (object) in Table~\ref{intersecting_Mbrane_WKK}.
In the Table, circles indicate where the brane world-volumes enter,
$\zeta$ represents the coordinate of the KK-monopole,
and the time-dependent branes are indicated by (a) and (b) and so on
for different solutions.
When the solutions can be used for cosmology and black hole physics,
they are marked in the corresponding columns.
The applications of these solutions to cosmology and black hole physics
are discussed in \S~\ref{Mwave_KKmonopole}.
\begin{table}[ht]
\caption{Intersecting M-branes with M-wave and KK-monopole.
Here we show only the interesting cases which can be applied to
cosmology or a black hole system.
The labelling (a), (b), $\cdots$ in the column ``$\tilde I$"
denotes which brane (or wave, KK-monopole) is time dependent.
In the second case of M2-M5-KKM system, there are two possibilities
which space dimensions can be our three space,
i.e., the case 1:
[$(\xi^1, \xi^2, \xi^3)
=(x^3, x^4, x^5)$] and the case 2:
[$(\xi^1, \xi^2, \xi^3)
=(x^7, x^8, x^9)$].
We show them by (d)-1
(d)-2, (e)-1, (e)-2, or (f)-1, (f)-2.
}
\label{intersecting_Mbrane_WKK}
\begin{center}
{\scriptsize
\begin{tabular}{|c||c|c|c|c|c|c|c|c|c|c|c|c||c|c|c|}
\hline
&&0&1&2&3&4&5&6&7&8&9&10&$\tilde{I}$&cos&BH\\
\hline
&M2 & $\circ$ & $\circ$ & $\circ$ &&&&&&&&&(a)&$-$&$\surd$ \\
\cline{3-13}
&M2 & $\circ$ &&& $\circ$ & $\circ$ &&&&&&&&&\\
\cline{3-13}
M2-M2-KKM&KKM & $\circ$ & $\circ$ & $\circ$ & $\circ$ & $\circ$ & $\circ$
& $\circ$ & $\zeta$ & ${\cal A}^{\rm (m)}_8$ & ${\cal A}^{\rm (m)}_9$
& ${\cal A}^{\rm (m)}_{10}$&(b)&$-$&$\surd$\\
\cline{2-16}
&M2 & $\circ$ & $\circ$ & $\circ$ &&&&&&&&&(c)&$\surd$&$-$ \\
\cline{3-13}
&M2 & $\circ$ &&& $\circ$ & $\circ$ &&&&&&&(d)&$\surd$&$-$\\
\cline{3-13}
&KKM & $\circ$ & $\zeta$ & ${\cal A}^{\rm (m)}_2$ & $\circ$ & $\circ$ & $\circ$
& $\circ$ &
$\circ$ & $\circ$ & ${\cal A}^{\rm (m)}_9$ & ${\cal A}^{\rm (m)}_{10}$
&(e)&$\surd$&$-$\\
\hline
\hline
&M2 & $\circ$ & $\circ$ & $\circ$ &&&&&&&&&(a)&$\surd$&$\surd$ \\
\cline{3-13}
M2-M5-W&M5 & $\circ$ & $\circ$ && $\circ$ & $\circ$ & $\circ$ & $\circ$ && & &
&(b)&$\surd$&$\surd$\\
\cline{3-13}
&W & $\circ$ & $\zeta$ &&& & & & &&&&(c)&$\surd$&$\surd$\\
\hline
&M2 & $\circ$ & $\circ$ & $\circ$ &&&&&&&&&(a)&$\surd$&$\surd$ \\
\cline{3-13}
&M5 & $\circ$ & $\circ$ && $\circ$ & $\circ$ & $\circ$ & $\circ$ &&& &
&(b)&$\surd$&$\surd$\\
\cline{3-13}
M2-M5-KKM
&KKM & $\circ$ & $\circ$ & $\circ$ & $\circ$ & $\circ$ & $\circ$ & $\circ$ &
$\zeta$ & ${\cal A}^{\rm (m)}_8$ & ${\cal A}^{\rm (m)}_9$
& ${\cal A}^{\rm (m)}_{10}$&(c)&$\surd$&$\surd$\\
\cline{2-16}
&M2 & $\circ$ & $\circ$ & $\circ$ &&&&&&&&&(d)-1, 2&$\surd$&$-$ \\
\cline{3-13}
&M5 & $\circ$ & $\circ$ && $\circ$ & $\circ$ & $\circ$ & $\circ$ &&& &
&(e)-1, 2&$\surd$&$-$\\
\cline{3-13}
&KKM & $\circ$ & $\zeta$ & ${\cal A}^{\rm (m)}_2$ & $\circ$ & $\circ$ & $\circ$
& ${\cal A}^{\rm (m)}_6$ &
$\circ$ & $\circ$ & $\circ$ & ${\cal A}^{\rm (m)}_{10}$&(f)-1, 2&$\surd$&$-$ \\
\hline
&M2 & $\circ$ & $\circ$ & $\circ$ &&&&&&&&&(a)&$\surd$&$\surd$ \\
\cline{3-13}
M2-M5-
&M5 & $\circ$ & $\circ$ && $\circ$ & $\circ$ & $\circ$ & $\circ$ &&& &
&(b)&$\surd$&$\surd$\\
\cline{3-13}
W-KKM
&W & $\circ$ & $\zeta^1$ &&& & & & &&&&(c)&$\surd$&$\surd$\\
\cline{3-13}
&KKM & $\circ$ & $\circ$ & $\circ$ & $\circ$ & $\circ$ & $\circ$ & $\circ$ &
$\zeta^7$ & ${\cal A}^{\rm (m)}_8$ & ${\cal A}^{\rm (m)}_9$
& ${\cal A}^{\rm (m)}_{10}$&(d)&$\surd$&$\surd$\\
\hline
\hline
&M5 & $\circ$ & $\circ$ & $\circ$ & $\circ$ & $\circ$ & $\circ$ &&&&&
&(a)&$-$&$\surd$ \\
\cline{3-13}
M5-M5-W
&M5 & $\circ$ & $\circ$ & $\circ$ & $\circ$ &&& $\circ$ & $\circ$ && &&&&\\
\cline{3-13}
&W & $\circ$ &$\zeta$ && & & & & &&&&(b)&$-$&$\surd$\\
\hline
&M5 & $\circ$ & $\circ$ & $\circ$ & $\circ$ & $\circ$ & $\circ$ &&&&&
&(a)&$\surd$&$-$ \\
\cline{3-13}
M5-M5-KKM
&M5 & $\circ$ & $\circ$ & $\circ$ & $\circ$ &&& $\circ$ & $\circ$ && &&
(b)&$\surd$&$-$\\
\cline{3-13}
&KKM & $\circ$ & $\circ$ & $\circ$ & $\circ$ & $\circ$ & $\circ$ & $\zeta$ &
${\cal A}^{\rm (m)}_7$ & $\circ$ & ${\cal A}^{\rm (m)}_9$
& ${\cal A}^{\rm (m)}_{10}$&(c)&$\surd$&$-$\\
\hline
&M5 & $\circ$ & $\circ$ & $\circ$ & $\circ$ & $\circ$ & $\circ$ &&&&&
&(a)&$\surd$&$-$ \\
\cline{3-13}
M5-M5-
&M5 & $\circ$ & $\circ$ & $\circ$ & $\circ$ &&& $\circ$ & $\circ$ && &
& & & \\
\cline{3-13}
KKM-KKM
&KKM & $\circ$ & $\circ$ & $\circ$ & $\circ$ & $\circ$ & $\circ$ & $\zeta^6$ &
${\cal A}^{\rm (m)}_7$ & $\circ$ & ${\cal A}^{\rm (m)}_9$
& ${\cal A}^{\rm (m)}_{10}$&(b)&$\surd$&$-$\\
\cline{3-13}
&KKM & $\circ$ & $\circ$ & $\circ$ & $\circ$ & $\zeta^4$ &
${\cal B}^{\rm (m)}_5$ & $\circ$ & $\circ$ & $\circ$ & ${\cal B}^{\rm (m)}_9$
& ${\cal B}^{\rm (m)}_{10}$&&&\\
\hline
\end{tabular}
}
\end{center}
\end{table}

\newpage

\end{document}